\documentclass[sigconf, nonacm, pdfa]{acmart}
\usepackage{balance}
\usepackage[a-2b]{pdfx}
\usepackage{amsfonts}
\usepackage{amsthm}
\usepackage{float}
\usepackage{placeins}
\usepackage{pgfplotstable}
\usepackage{pgfplots}
\usepackage{scalefnt}
\usepackage{tikz}
\usetikzlibrary{patterns}
\usetikzlibrary{pgfplots.groupplots}
\usepgfplotslibrary{groupplots}
\usepgfplotslibrary{statistics}
\pgfplotsset{
    table/search path={.//data/}
}
\usepackage[ruled,vlined]{algorithm2e}
\usepackage{algpseudocode}
\usepackage{subcaption}
\usepackage{lscape}
\usepackage{tabularx}
\usepackage{hyperref}
\usepackage{cleveref}
\usepackage{booktabs}
\usepackage{graphicx}
\usepackage{xcolor}
\usepackage{enumitem}

\usepackage{amsmath}
\usepackage[colorinlistoftodos]{todonotes}
\usepackage{indentfirst}
\usepackage{float}
\usepackage{threeparttable} 
\usepackage{multirow}
\usepackage{url}
\usepackage{flushend}
\usepackage[all]{nowidow}
\usepackage[most]{tcolorbox}

\newcolumntype{b}{>{\hsize=2.3\hsize}X}
\newcolumntype{s}{>{\hsize=.45\hsize}X}
\newcolumntype{m}{>{\hsize=.9\hsize}X}

\definecolor{airforceblue}{rgb}{0.36, 0.54, 0.66}
\definecolor{aliceblue}{rgb}{0.94, 0.97, 1.0}
\definecolor{alizarin}{rgb}{0.82, 0.1, 0.26}
\definecolor{amber}{rgb}{1.0, 0.75, 0.0}
\definecolor{americanrose}{rgb}{1.0, 0.01, 0.24}
\definecolor{azure(colorwheel)}{rgb}{0.0, 0.5, 1.0}
\definecolor{amethyst}{rgb}{0.6, 0.4, 0.8}
\definecolor{ballblue}{rgb}{0.13, 0.67, 0.8}
\definecolor{babyblue}{rgb}{0.54, 0.81, 0.94}
\definecolor{brightgreen}{rgb}{0.4, 1.0, 0.0}
\definecolor{brightturquoise}{rgb}{0.03, 0.91, 0.87}
\definecolor{cadmiumyellow}{rgb}{1.0, 0.96, 0.0}
\definecolor{caribbeangreen}{rgb}{0.0, 0.8, 0.6}
\definecolor{celadon}{rgb}{0.67, 0.88, 0.69}
\definecolor{darkpastelgreen}{rgb}{0.01, 0.75, 0.24}
\definecolor{darktangerine}{rgb}{1.0, 0.66, 0.07}
\definecolor{darkturquoise}{rgb}{0.0, 0.81, 0.82}
\definecolor{electriccyan}{rgb}{0.0, 1.0, 1.0}
\definecolor{ferrarired}{rgb}{1.0, 0.11, 0.0}
\definecolor{electricblue}{rgb}{0.49, 0.98, 1.0}
\definecolor{electriclime}{rgb}{0.8, 1.0, 0.0}
\definecolor{lasallegreen}{rgb}{0.03, 0.47, 0.19}

\newcommand{\graphname}{RoarGraph\xspace}

\newcommand{\revisionred}[1]{\textcolor{blue}{#1}}
\renewcommand{\revisionred}[1]{#1}

\newcommand{\finalred}[1]{\textcolor{red}{#1}}
\renewcommand{\finalred}[1]{#1}

\newcommand\vldbdoi{10.14778/3681954.3681959}
\newcommand\vldbpages{2735 - 2749}
\newcommand\vldbvolume{17}
\newcommand\vldbissue{11}
\newcommand\vldbyear{2024}
\newcommand\vldbauthors{\authors}
\newcommand\vldbtitle{\shorttitle} 
\newcommand\vldbavailabilityurl{https://github.com/matchyc/RoarGraph}
\newcommand\vldbpagestyle{empty} 

\begin{document}



\title{RoarGraph: A Projected Bipartite Graph for Efficient Cross-Modal Approximate Nearest Neighbor Search}



\author{Meng Chen}
\affiliation{%
  \institution{Fudan University}
}
\email{mengchen22@m.fudan.edu.cn}

\author{Kai Zhang}
\affiliation{%
  \institution{Fudan University}
}
\email{zhangk@fudan.edu.cn}

\author{Zhenying He}
\affiliation{%
  \institution{Fudan University}
}
\email{zhenying@fudan.edu.cn}

\author{Yinan Jing}
\affiliation{%
  \institution{Fudan University}
}
\email{jingyn@fudan.edu.cn}

\author{X.Sean Wang}
\affiliation{%
  \institution{Fudan University}
}
\email{xywangcs@fudan.edu.cn}

\begin{abstract}

Approximate Nearest Neighbor Search (ANNS) is a fundamental and critical component in many applications, including recommendation systems and large language model-based applications. With the advancement of multimodal neural models, which transform data from different modalities into a shared high-dimensional space as feature vectors, cross-modal ANNS aims to use the data vector from one modality (e.g., texts) as the query to retrieve the most similar items from another (e.g., images or videos). However, there is an inherent distribution gap between embeddings from different modalities, and cross-modal queries become Out-of-Distribution (OOD) to the base data. Consequently, state-of-the-art ANNS approaches suffer poor performance for OOD workloads.

In this paper, we quantitatively analyze the properties of the OOD workloads to gain an understanding of their ANNS efficiency. Unlike single-modal workloads, we reveal OOD queries spatially deviate from base data, and the k-nearest neighbors of an OOD query are distant from each other in the embedding space. The property breaks the assumptions of existing ANNS approaches and mismatches their design for efficient search. With the insights from the OOD workloads, we propose p\textbf{Ro}jected bip\textbf{ar}tite \textbf{Graph} (\textbf{\graphname}), an efficient ANNS graph index that is built under the guidance of query distribution. Extensive experiments show that \graphname significantly outperforms state-of-the-art approaches on modern cross-modal datasets, achieving up to 3.56$\times$ faster search speed at a 90\% recall rate for OOD queries.

\end{abstract}

\maketitle

\pagestyle{\vldbpagestyle}
\begingroup\small\noindent\raggedright\textbf{PVLDB Reference Format:}\\
\vldbauthors. \vldbtitle. PVLDB, \vldbvolume(\vldbissue): \vldbpages, \vldbyear.\\
\href{https://doi.org/\vldbdoi}{doi:\vldbdoi}
\endgroup
\begingroup
\renewcommand\thefootnote{}\footnote{\noindent
The corresponding author is Dr. Kai Zhang. \\
This work is licensed under the Creative Commons BY-NC-ND 4.0 International License. Visit \url{https://creativecommons.org/licenses/by-nc-nd/4.0/} to view a copy of this license. For any use beyond those covered by this license, obtain permission by emailing \href{mailto:info@vldb.org}{info@vldb.org}. Copyright is held by the owner/author(s). Publication rights licensed to the VLDB Endowment. \\
\raggedright Proceedings of the VLDB Endowment, Vol. \vldbvolume, No. \vldbissue\ %
ISSN 2150-8097. \\
\href{https://doi.org/\vldbdoi}{doi:\vldbdoi} \\
}\addtocounter{footnote}{-1}\endgroup

\ifdefempty{\vldbavailabilityurl}{}{
\vspace{.3cm}
\begingroup\small\noindent\raggedright\textbf{PVLDB Artifact Availability:}\\
The source code, data, and/or other artifacts have been made available at \url{\vldbavailabilityurl}.
\endgroup
}

\section{Introduction}


\label{sec1}
Approximate nearest neighbor search (ANNS) is a fundamental and performance-critical component in various application domains such as large-scale information retrieval \cite{DBLP:conf/iclr/XiongXLTLBAO21,nigam2019semantic,long2022retrieval}, recommendation \cite{pal2020pinnersage,DBLP:conf/recsys/CovingtonAS16}, and question answering \cite{seo2019real,DBLP:journals/tacl/LewisWLMKPSR21}. More recent applications of retrieval-augmented generation (RAG) in large language models (LLMs) also utilize vector databases as external knowledge libraries, employing ANNS to enhance search efficiency \cite{liu2023learning,wang2023instructretro,asai-etal-2023-retrieval}. These applications demand \finalred{fast and accurate responses} to similarity vector search, where ANNS can be performed to efficiently retrieve the approximate nearest neighbors from the database for a given query, rather than conducting impractically exact k-nearest neighbor searches \cite{zhao2020song,chen2023finger,DBLP:journals/pami/FuWC22}. To improve the ANNS performance, a spectrum of studies have been carried out to design efficient data structures, including partition-based approaches \cite{DBLP:conf/cvpr/Silpa-AnanH08,DBLP:conf/soda/Yianilos93, DBLP:journals/corr/DolatshahHM15,sivic2003video}, quantization-based methods \cite{DBLP:conf/cvpr/BabenkoL14,DBLP:conf/cvpr/GeHK013,jegou2010product,DBLP:conf/nips/WuGSKHSY17,guo2020accelerating}, and hashing-based methods \cite{DBLP:conf/compgeom/DatarIIM04,DBLP:conf/sigmod/GanFFN12,DBLP:journals/pvldb/HuangFZFN15,DBLP:journals/pvldb/HuangFZFN15,DBLP:journals/vldb/ZhengZWNLJ22}, where graph-based approaches \cite{DBLP:journals/pami/MalkovY20,DBLP:journals/pvldb/FuXWC19,DBLP:journals/pacmmod/PengCCYX23} represent the state-of-the-art performance on many datasets.




Recently, \finalred{cross-modal retrieval has drawn much attention with the advancement of multimodal data representation techniques. Deep learning models, such as CLIP \cite{DBLP:conf/icml/RadfordKHRGASAM21}}, trained for multimodal tasks embed unstructured data from different modalities like vision and natural language into a shared high-dimensional space with semantics preserved, say embeddings.
In cross-modal vector search, data from one modal (e.g., texts) is used as the query to retrieve the most semantically similar data from another modal (e.g., images or videos) \cite{liu2021hit,DBLP:conf/cvpr/HuangGPJLLW23,lei2021less}. With the diverse and critical application scenarios, \finalred{efficient ANNS are widely demanded to enhance the performance of cross-modal retrieval} \cite{pmlr-v176-simhadri22a,DBLP:journals/corr/WangY0W016,DBLP:conf/sigir/YuFL22,yang2022mutual,DBLP:journals/pvldb/SunWQZL14,DBLP:conf/iclr/SheyninAPSGNT23,cao2016correlation}. However, existing ANNS indexes are designed for single-modal scenarios, and they suffer poor performance with cross-modal queries. For example, in the modern cross-modal dataset LAION \cite{schuhmann2021laion}, an HNSW \revisionred{(Hierarchical Navigable Small World) \cite{DBLP:journals/pami/MalkovY20}} index needs to visit 14374 nodes to ensure recall@10 to reach 0.95 in text-image search, while only 1568 nodes are required to traverse if using an image to search images, indicating nearly 10 times efficiency degradation.


The main characteristic of cross-modal retrieval is the dramatically different data distributions between vectors from two modalities. Even though multimodal neural embedding models enable similarity measurements on vectors from different modalities, a consistent and inherent distribution gap, recognized as the \textit{modality gap}, persists between embeddings of two modalities in cross-modal representation learning \cite{liang2022mind}. Accordingly, in cross-modal ANNS, query vectors are \textit{Out-of-Distribution (OOD)} with respect to vectors in the database (base data) \cite{DBLP:journals/corr/abs-2211-12850}. This stands in stark contrast to single-modal workloads where queries are \textit{In-Distribution (ID)} with the base data. The Mahalanobis distance \cite{mahalanobis2018generalized} shows that queries from another modality deviate 10 $\sim$ 100$\times$ far away from the base data than that between ID queries and the base data in cross-modal datasets, e.g. Text-to-Image \cite{yandextexttoimage}, LAION \cite{schuhmann2021laion}, and WebVid \cite{Bain21}. Further, through in-depth experiments and analysis, we find that \textbf{an out-of-distribution query is far from the base data}, and \textbf{the k-nearest neighbors of such a query tend to be distant from each other}, indicating that queries deviate from the base data and the nearest neighbors (ground truths) \finalred{to OOD queries are more widely distributed than that to ID queries.}


However, state-of-the-art ANNS indexes are designed for ID queries \cite{DBLP:journals/pvldb/FuXWC19,DBLP:journals/pami/MalkovY20,DBLP:journals/pacmmod/PengCCYX23}. They presume that queries appear near the base data and nearest neighbors for a query are in close proximity to each other.
Under the assumption, graph-based ANNS approaches employ beam search (n-greedy search) during the index-building phase to construct an approximate KNN graph \cite{DBLP:conf/sigmod/LiZAH20,DBLP:journals/pvldb/LuKXI21,DBLP:journals/pami/MalkovY20,DBLP:journals/pvldb/FuXWC19,jayaram2019diskann}, where vectors with a smaller distance tend to be connected.
Besides, the search phase also uses beam search, which anticipates rapid convergence with the closely connected base data. This indexing convention suffers from OOD queries in cross-modal ANNS. \finalred{Since cross-modal queries and the base data follow different distributions and the ground truths for OOD queries are scattered}, \textbf{the critical assumption on the distribution of queries and the base data is broken}. 
As a result, the cross-modal search on such a graph cannot converge efficiently but incurs more hops in graph traversals.
That constitutes the primary reason why existing ANNS approaches are incompetent in handling OOD workloads.


We propose p\textbf{Ro}jected bip\textbf{ar}tite \textbf{Graph} (\textbf{\graphname}), an efficient graph index with the knowledge from query distribution for cross-modal ANNS. \finalred{Our key idea is to map distributed vectors that are nearest neighbors to queries into closely connected neighbors within a graph index}. The indexing process of \graphname is as follows.
Firstly, with elaborate edge selection, a bipartite graph is built by mapping the relationship of similarity between queries and base data into the unified graph structure. Secondly, the bipartite graph is projected onto base data, incorporating neighborhood-aware projection to create pathways for spatially distant nodes, recognized as proximate from the perspective of queries. Finally, a connectivity enhancement scheme is performed to ensure the graph's connectivity and the reachability of all base data vectors. The \graphname index exclusively consists of base data yet effectively preserves the neighboring relationship derived from the query distribution.
The main contributions of this paper are summarized as follows:
\begin{itemize}

    \item We identify the inefficiency of cross-modal ANNS through in-depth experiments and present an insightful analysis that reveals underlying reasons causing performance degradation on state-of-the-art approaches in cross-modal ANNS.
    \item We propose \finalred{\graphname, a novel graph index for efficient cross-modal ANNS}, which effectively utilizes query distribution to guide graph index construction. 
    \item We performed extensive experiments on three cross-modal datasets comprising text, images, and video frames. Our results show that \graphname significantly improves the performance of cross-modal vector search.
\end{itemize}

\graphname speeds up cross-modal vector search on a graph by minimizing detours and reducing the number of hops during the search phase. \finalred{This leads to a significant performance improvement over existing graph indexes}, ranging from 1.84$\times$ to 3.56$\times$ faster on three cross-modal datasets with recall@k$\ge$0.9, where \finalred{k=1,10, and 100}. In particular, \graphname also achieves an exceptional level of recall (recall@k $\ge$ 0.99) that is unattainable by existing methods. A variant of our approach won the championship in the OOD track of NeurIPS' Practical Vector Search (Big ANN) Challenge 2023.

\section{Background and Motivation}
\label{sec:section-2}
This section introduces the background of approximate nearest neighbor search (ANNS) and out-of-distribution (OOD) ANNS. We also quantitatively evaluate and analyze the performance of existing ANNS approaches on cross-modal datasets.



\subsection{Background on ANNS}

\subsubsection{The ANNS Definition}
ANNS stems from the k-nearest neighbors search (KNNS), which aims to find $k$ vectors in the database (base data) that are closest to a given query vector. The measurement of closeness typically involves utilizing cosine distance, $\ell_2$-distance, inner product, etc. Contemporary applications involve working with large-scale datasets in which the dimension of vectors grows to hundreds \cite{pmlr-v176-simhadri22a}, the pursuit of the exact KNNS becomes both costly and impractical due to the challenges imposed by the \textit{curse of dimensionality} \cite{DBLP:conf/stoc/IndykM98}. ANNS approaches create specific indexes for the base data to achieve a tradeoff between search speed and accuracy.

\newtheorem{ann_def1}{Definition}

\begin{ann_def1}
Given $N$ vectors in the database $\mathcal{X} = \{x_1,...,x_N\} \in \mathbb{R}^D$, $q \in \mathbb{R}^D$ and a function for computing the distance between two vectors $\delta(\cdot, \cdot)$. Top-k ( $k \le N$) ANNS amis to find
 \begin{equation}
    S = \underset{i\in{1,...,N}}{\mathrm{k\text{-}\arg min}}\ \delta(q, x_i)
 \end{equation}
$S$ satisfies that $|S| = k$, $\delta(q, x) \le (1 + \epsilon)\delta({x}', q)$ for $x \in S, {x}' \in \mathcal{X} \setminus S$ and $\epsilon \ge 0$, where $\epsilon$ is a small constant and not used directly, only denoting the approximation property hold by ANNS.
\end{ann_def1}

\label{sec:section-2-1}
The search performance of an ANN index will be evaluated by search speed-vs-recall tradeoff. Recall is calculated using the formula recall@$k=|S \cap KNN(q)|/k$, where $KNN(q)$ represents the exact k-nearest neighbors (ground truths) of a query $q$, and $S$ is the result set with $|S|=k$.


\subsubsection{State-of-the-Art ANNS Approaches}



\textbf{Graph-based methods} and the \textbf{inverted file index} are two prevalent ANNS approaches.
Graph-based methods are the most high-performance ANNS approach family, providing a better search speed-recall tradeoff than other methods \cite{DBLP:journals/pami/MalkovY20,DBLP:journals/pvldb/FuXWC19,DBLP:journals/pacmmod/PengCCYX23,DBLP:journals/tkde/LiZSWLZL20,DBLP:journals/pvldb/WangXY021,DBLP:journals/pvldb/LuKXI21}. These methods index base data into a graph structure that each node represents one data vector. HNSW \cite{DBLP:journals/pami/MalkovY20} is a well-known multi-layer graph in the hierarchical structure. During construction, all base vectors are inserted into the base layer (layer 0), and only subsets of layer $i$ will be inserted into the $i+1$ layer, with decreasing probability from bottom to top (the top layer contains only 1 point). Nodes within each layer are connected to their approximate nearest neighbors. During the search phase, HNSW employs greedy search on higher layers for a given query. The closest point obtained from higher layers becomes the entry point for the next layer. At the base layer, a beam search is performed on the graph. Beam search is a variant of greedy search that explores a graph by expanding the most promising element in a limited queue \cite{DBLP:conf/icml/ProkhorenkovaS20}, converging towards the nearest neighbors for a given query. The capacity of the queue, termed the beam width, controls the trade-off between accuracy and search speed.

\begin{table}[t]
\centering
\caption{Cross-modal Datasets}
\begin{tabularx}{\linewidth}{
  | c 
  | c
  | c
  | >{\centering\arraybackslash}X
  | >{\centering\arraybackslash}X| 
}
\hline
\textbf{Dataset} &
\textbf{Scale} &
  \textbf{\begin{tabular}[c]{@{}c@{}}Vector \\ Dimension\end{tabular}} &
  \textbf{\begin{tabular}[c]{@{}c@{}}Type of\\ Query\end{tabular}} &
  \textbf{\begin{tabular}[c]{@{}c@{}}Type of\\ Base\end{tabular}} \\ \hline
Text-to-Image\cite{yandextexttoimage} & 10M & 200 & Text & Image \\ \hline
LAION\cite{schuhmann2021laion} & 10M         & 512 & Text & Image \\ \hline
WebVid\cite{Bain21} & 2.5M & 512& Text & Video \\ \hline
\end{tabularx}
\label{table:datasets}
\end{table}
\begin{table}[t]
\centering
\caption{2-Wasserstein Distances on Cross-modal Datasets}
\begin{tabularx}{\linewidth}{
  | c 
  | c
  | >{\centering\arraybackslash}X
  | >{\centering\arraybackslash}X| 
}
\hline
\textbf{Dataset} & \textbf{$W_2(B_1, B_2)$ }& \textbf{$W_2(B_1, Q)$} & \textbf{$W_2(B_2, Q)$} \\ \hline
Text-to-Image & 0.864 & 1.439 & 1.437 \\ \hline
LAION & 0.882 & 1.677 & 1.677 \\ \hline
WebVid & 0.581 & 1.683 & 1.674 \\ \hline
\end{tabularx}
\label{tab:w2dist}
\end{table}
The inverted file index (IVF)-based methods are another popular index type in ANNS for its convenience and superior performance for range nearest neighbors search, including IVF \cite{sivic2003video}, IMI \cite{babenko2014inverted}, etc. IVF first applies \finalred{K-means} to \textit{cluster} base data and gets $n$ centroids, then assigns each vector in base data to its nearest cluster. During the search phase, IVF selects $nprobe$ closest centroids to the query, then scans all vectors belonging to the corresponding $nprobe$ clusters and obtains the top-k results.

\subsection{Out-of-Distribution ANNS}
Out-of-distribution Approximate Nearest Neighbor Search (OOD-ANNS) \finalred{indicates the distribution of queries} differs from that of the base data. A distribution gap is consistently present and inherent between vectors from two modalities in modern cross-modal applications \cite{liang2022mind}. For instance, when using texts as queries to retrieve relevant visual data, a distribution gap occurs, resulting in queries becoming out-of-distribution. Unfortunately, the majority of ANNS indexing algorithms are primarily designed for single-modal tasks, such as image-image search.


The OOD property of cross-modal vector search can be mathematically quantified using two mathematical distances: the Wasserstein distance \cite{vaserstein1969markov, kantorovich1960mathematical} that measures two distributions and the Mahalanobis distance \cite{mahalanobis2018generalized} that measures the distance from a vector to a distribution. \finalred{We use two metrics to evaluate data distributions on three modern real-world multimodal datasets: Text-to-Image \cite{yandextexttoimage}, LAION \cite{schuhmann2021laion}, and WebVid \cite{Bain21}. The characteristics of these datasets are shown in \Cref{table:datasets}.}


To quantify the OOD characteristic by Wasserstein distance, we sample two non-intersecting sets ($B_1$, $B_2$) from the base data and one query set ($Q$) from the query vectors, each containing 100,000 vectors. As shown in \Cref{tab:w2dist}, two samples from the base data demonstrate proximity with $W_2(B_1,B_2)$. In contrast, the query distribution diverged from the base data distribution, being 1.67 times, 1.89 times, and 2.89 times more distant across three datasets.

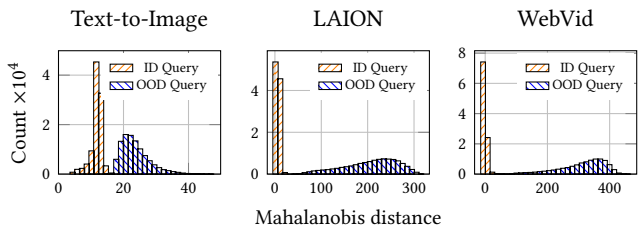
\begin{figure}[!t]
    \centering
    \pgfplotsset{
                width=0.21\textwidth,
                legend image post style={scale=0.5},
            legend style={
                draw=none, 
                nodes={scale=0.6, transform shape},
                anchor=north,
                at={(0.625, 0.98)},
                legend columns=1,
                },
            }
        \begin{tikzpicture}[baseline]
        \begin{axis}[
            y filter/.expression={y==0 ? nan : y},
            ylabel={Count $\times 10^{4}$},
            ymin=0,
            xmin=0,
            bar width=2,
            ybar,
            area legend,
            ymajorgrids=true, 
            xmajorgrids=true, 
                    ylabel near ticks,
        xlabel near ticks,
        xtick align=inside,
            xticklabel style={font=\scriptsize}, 
            yticklabel style={font=\scriptsize}, 
                        label style={font=\small},
    title={Text-to-Image},
    title style={yshift=-0.5ex},
                    y filter/.code={\pgfmathdivide{#1}{10000}},
        ]
        \addplot  [pattern=north east lines, pattern color=orange] 
            table[col sep=comma, x=x, y=id_y] {t2i-mdist.dat}; 
        
        \addplot [pattern=north west lines, pattern color=blue] 
            table[col sep=comma, x=x, y=ood_y] {t2i-mdist.dat};
        \legend{ID Query, OOD Query};
        \end{axis}
\end{tikzpicture}\hfill%
    \label{fig:t2i-mdist}
    \begin{tikzpicture}[baseline]
        \begin{axis}[
            y filter/.expression={y==0 ? nan : y},
            ymin=0,
            bar width=2,
            ybar,
            xlabel=Mahalanobis distance,
            area legend,
            ymajorgrids=true, 
            xmajorgrids=true, 
            xtick style={draw=none}, 
        ylabel near ticks,
        xlabel near ticks,
        xtick align=inside,
        xticklabel style={font=\scriptsize}, 
        yticklabel style={font=\scriptsize}, 
        label style={font=\small},
        title={LAION},
                        y filter/.code={\pgfmathdivide{#1}{10000}},
        ]
        \addplot  [pattern=north east lines, pattern color=orange] 
            table[col sep=comma, x=x, y=id_y] {laion-mdist.dat};
        
        \addplot [pattern=north west lines, pattern color=blue] 
            table[col sep=comma, x=x, y=ood_y] {laion-mdist.dat};
        \legend{ID Query, OOD Query};
        \end{axis}
    \end{tikzpicture}\hspace{-0.07cm}%
    \label{fig:laion-m-dist}
    \begin{tikzpicture}[baseline]
        \begin{axis}[
            y filter/.expression={y==0 ? nan : y},
            ymin=0,
            bar width=1.8,
            ybar,
            area legend,
            ymajorgrids=true, 
            xmajorgrids=true, 
            xtick style={draw=none}, 
                    ylabel near ticks,
        xlabel near ticks,
        xtick align=inside,
                    xticklabel style={font=\scriptsize}, 
            yticklabel style={font=\scriptsize}, 
                        label style={font=\small},
                            title={WebVid},
                    y filter/.code={\pgfmathdivide{#1}{10000}},
        ]
        \addplot  [pattern=north east lines, pattern color=orange] 
            table[col sep=comma, x=x, y=id_y] {webvid-mdist.dat};
        
        \addplot [pattern=north west lines, pattern color=blue] 
            table[col sep=comma, x=x, y=ood_y] {webvid-mdist.dat};
        \legend{ID Query, OOD Query};
        \end{axis}
    \end{tikzpicture}%
    \label{fig:webvid-m-dist}
\caption{\finalred{Mahalanobis distances from OOD/ID queries to the base data.}}

\label{fig:mdist-hist}
\end{figure}
In addition to the entire distribution difference, a query is considered out-of-distribution if its Mahalanobis distance to the base data significantly differs from the distances between base vectors \cite{DBLP:journals/corr/abs-2211-12850}. For each $q_{id}$ in the ID query set and $q_{ood} \in Q$, we compute the Mahalanobis distance to estimate $d_{M}(q, P)$, where $P$ is the base data distribution. As depicted in \Cref{fig:mdist-hist}, it is evident that OOD queries significantly deviate from the base data distribution. \finalred{In particular, queries from LAION and WebVid exhibit a more pronounced out-of-distribution property compared to those in Text-to-Image.}

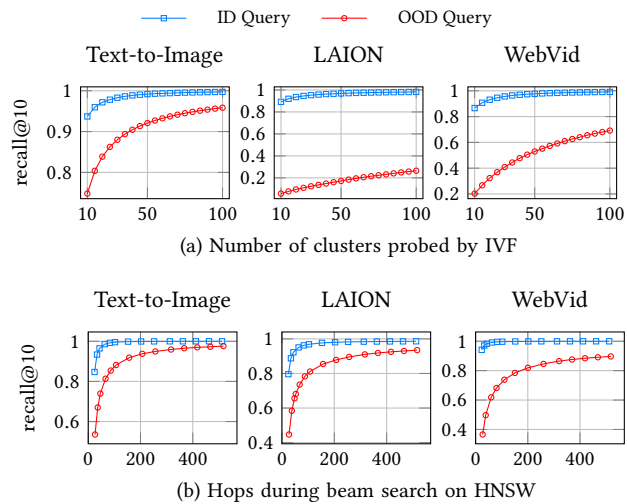
\begin{figure}[t]
   \centering
   \ref{m-performance-grouplegend}
  \pgfplotsset{
    compat=1.14,
        ylabel near ticks,
       xlabel near ticks,
            legend style={
                draw=none, 
                nodes={scale=0.8, transform shape},
                },
            }
\begin{tikzpicture}[baseline]
        \begin{groupplot}[
group style = {
group size = 3 by 1,
horizontal sep=17,
ylabels at=edge left,
},
width = 0.2\textwidth,
       xlabel near ticks,
       grid=major,
       yticklabel style={font=\small},
       xticklabel style={font=\small},
       xmin=10,
       xmax=100,
       enlargelimits=0.05,
       xlabel style ={font=\small, text width=2.8cm, align=center},
       xtick={10, 50, 100},
]
\nextgroupplot[ 
            title = {Text-to-Image},
            title style={yshift=-0.5ex},
            legend style = { 
            column sep = 10pt, 
            legend columns = -1, 
            },
            xticklabel style={font=\small},
            ylabel={recall@10},
            ylabel style={
            font=\small,
            text width=2cm,
            align=center},
            ]
xticklabel style={font=\tiny},

\addplot[color=azure(colorwheel),mark=square,  mark size=1,] table [x=nprobe, y=recall, col sep=comma] {bench_data//faiss-ivf-t2i10M-10240-r10-id-bench.dat};
\addplot[color=red,mark=o, mark size=1] table [x=nprobe, y=recall, col sep=comma] {bench_data//faiss-ivf-t2i10M-10240-r10-bench.dat};

\nextgroupplot[
        title = LAION,
        xlabel=(a) Number of clusters probed by IVF,
        xlabel style = {
        font=\small,
        text width=5cm,
        }
]

\addplot[color=azure(colorwheel),mark=square,  mark size=1,] table [x=nprobe, y=recall, col sep=comma] {bench_data//faiss-ivf-laion10M-10240-r10-id-bench.dat};
\addplot[color=red,mark=o, mark size=1] table [x=nprobe, y=recall, col sep=comma] {bench_data//faiss-ivf-laion10M-10240-r10-bench.dat};

\nextgroupplot[
        title=WebVid,
        yticklabel style={font=\small},
        xticklabel style={font=\small},
]

\addplot[color=azure(colorwheel),mark=square,  mark size=1,] table [x=nprobe, y=recall, col sep=comma] {bench_data//faiss-ivf-webvid1M-10240-r10-id-bench.dat};
\addplot[color=red,mark=o, mark size=1] table [x=nprobe, y=recall, col sep=comma] {bench_data//faiss-ivf-webvid1M-10240-r10-bench.dat};

\end{groupplot}
\end{tikzpicture}%

\begin{tikzpicture}[baseline]
\begin{groupplot}[
group style = {
group size = 3 by 1,
horizontal sep=17,
ylabels at=edge left,
},
width = 0.2\textwidth,
       xlabel near ticks,
       grid=major,
       yticklabel style={font=\small},
       xticklabel style={font=\small},
       xlabel style ={font=\small, text width=2.8cm, align=center},
       xmin=0,
]
\nextgroupplot[ 
            title = {Text-to-Image},
            title style={yshift=-0.5ex},
            legend style = { 
            column sep = 10pt, 
            legend columns = -1, 
            },
            legend to name = m-performance-grouplegend,
            xticklabel style={font=\small},
            ylabel={recall@10},
            ylabel style={
            font=\small,
            text width=3cm,
            align=center},
            ]
           xticklabel style={font=\small},

\addlegendentry{ID Query}
\addplot[
color=azure(colorwheel),
mark=square,  mark size=1, 
] 
table [x=hops, y=recall, col sep=comma] {bench_data//hnsw-t2i10M-M32-r10-id-bench.dat};
\addlegendentry{OOD Query}
\addplot[color=red,
mark=o, mark size=1] table [x=hops, y=recall, col sep=comma] {bench_data//hnsw-t2i10M-M32-r10-bench.dat};
\nextgroupplot[
        title = LAION,
        xlabel={(b) Hops during beam search on HNSW},
        xlabel style = {
            font=\small,
            text width=5cm,
        },
]

\addplot[color=azure(colorwheel),
mark=square,  mark size=1,] table [x=hops, y=recall, col sep=comma] {bench_data//hnsw-laion10M-M32-r10-id-bench.dat};
\addplot[color=red,
mark=o, mark size=1] table [x=hops, y=recall, col sep=comma] {bench_data//hnsw-laion10M-M32-r10-bench.dat};

\nextgroupplot[
        title=WebVid,
        yticklabel style={font=\small},
        xticklabel style={font=\small},
]

\addplot[color=azure(colorwheel),
mark=square,  mark size=1,] table [y=recall, x=hops, col sep=comma] {bench_data//hnsw-webvid1M-M32-r10-id-bench.dat};
\addplot[color=red,
mark=o, mark size=1] table [y=recall, x=hops, col sep=comma] {bench_data//hnsw-webvid1M-M32-r10-bench.dat};

\end{groupplot}
    \end{tikzpicture}

    \caption{Performance evaluation on ID and OOD workloads.}
    \label{fig:motivation-evaluation}
\end{figure}

\subsection{The Inefficiency of Current Approaches for OOD-ANNS}


\subsubsection{Performance of Current Approaches on OOD Workloads}
\label{sec:section-2-3-1-ood-eval}


Evaluations of IVF and HNSW are conducted on three multimodal datasets shown in \Cref{table:datasets}. In-distribution (ID) queries are sourced from original large-scale datasets that follow the same empirical distribution of base data, and OOD queries use the textual query set for each dataset. We build IVF indexes by the Faiss library \cite{johnson2019billion} and HNSW indexes by the official implementations \cite{DBLP:journals/pami/MalkovY20} with recommended parameters.

\label{sec:2-3-1-scann}
Critical performance degrading on OOD workloads is observed in \Cref{fig:motivation-evaluation}. When using the IVF index, OOD queries demonstrate a noticeable need to \finalred{search a significantly larger number of clusters} to achieve high recalls, in contrast to ID queries. Recall@10 for ID queries exceeded 0.97 when searching the closest 50 clusters on all three datasets, whereas OOD queries achieved recall@10 of only 0.91, 0.20, and 0.52, respectively.

Similarly, with the HNSW index, OOD queries also require visiting a much larger number of nodes during beam search on the graph, resulting in poor search efficiency among the three datasets. On the LAION dataset, OOD queries necessitate more than 500 hops to achieve recall@10$\ge$0.93, whereas only 48 hops are required for ID queries. This highlights inefficient performance attributed to approximately 10 times the length of the search path caused by OOD queries. The substantial performance decay proves that existing indexes perform poorly on cross-modal ANNS tasks, emphasizing the urgent need for an efficient index.



\begin{figure}[t]
    \centering
    \includegraphics[width=\linewidth]{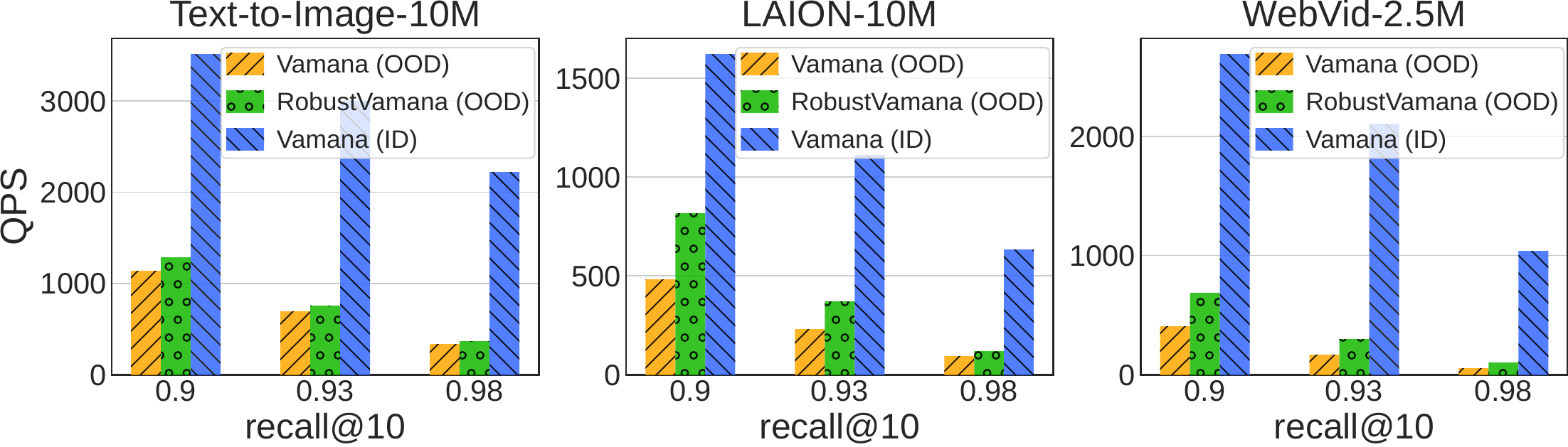}
    
    \caption{Evaluation of OOD-DiskANN and DiskANN. The notations ID and OOD in parentheses denote ID and OOD query workloads, respectively.}
    \label{fig:ood-diskann-improvement}

\end{figure}
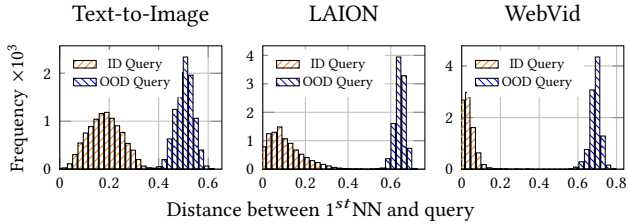
\begin{figure}[!t]
    \centering
    \pgfplotsset{
    width=0.21\textwidth,
    legend pos=north west,
  legend image post style={scale=0.5},
            legend style={
                draw=none, 
                nodes={scale=0.6, transform shape},
                legend columns=1,
                },
        }

        \begin{tikzpicture}[baseline]
        \begin{axis}[
            title=Text-to-Image,
            title style={yshift=-0.5ex},
            y filter/.expression={y==0 ? nan : y},
            xmin=0,
            ymin=0,
            xtick distance=0.2,
            y filter/.code={\pgfmathdivide{#1}{1000}},
            bar width=1.68,
            ybar,
            area legend,
            ymajorgrids=true, 
            xmajorgrids=true, 
            xticklabel style={font=\scriptsize}, 
            yticklabel style={font=\scriptsize}, 
                    ylabel near ticks,
        xlabel near ticks,
        xtick align=inside,
            ylabel={Frequency $\times 10^{3}$},
            ylabel style={
            font=\small,
            text width=2cm,
            align=center, scale=0.9},
        ]
        \addplot  [pattern=north east lines, pattern color=orange] 
            table[col sep=comma, x=x, y=id_y] {t2i10M-q10k-q-gt-dist.dat};
        
        \addplot [pattern=north west lines, pattern color=blue] 
            table[col sep=comma, x=x, y=ood_y] {t2i10M-q10k-q-gt-dist.dat};
        \legend{ID Query, OOD Query};
        \end{axis}
    \end{tikzpicture}
    \label{fig:t2i10M-q-gt-dist}
        \begin{tikzpicture}[baseline]
        \begin{axis}[
            title=LAION,
            y filter/.expression={y==0 ? nan : y},
            y filter/.code={\pgfmathdivide{#1}{1000}},
            ymin=0,
            xmin=0,
            bar width=1.68,
            ybar,
            area legend,
            ymajorgrids=true, 
            xmajorgrids=true, 
            xtick style={draw=none}, 
            xticklabel style={font=\scriptsize}, 
            yticklabel style={font=\scriptsize}, 
                    ylabel near ticks,
        xlabel near ticks,
        xtick align=inside,
        ]
        \addplot  [pattern=north east lines, pattern color=orange] 
            table[col sep=comma, x=x, y=id_y] {laion10M-q10k-q-gt-dist.dat};
        
        \addplot [pattern=north west lines, pattern color=blue] 
            table[col sep=comma, x=x, y=ood_y] {laion10M-q10k-q-gt-dist.dat};
        \legend{ID Query, OOD Query};
        \end{axis}
    \end{tikzpicture}
    \label{fig:laion10M-q-gt-dist}
        \begin{tikzpicture}[baseline]
        \begin{axis}[
            title=WebVid,
            y filter/.expression={y==0 ? nan : y},
            y filter/.code={\pgfmathdivide{#1}{1000}},
            ymin=0,
            xmin=0,
            bar width=1.68,
            ybar,
            area legend,
            ymajorgrids=true, 
            xmajorgrids=true, 
            xtick style={draw=none}, 
            xticklabel style={font=\scriptsize}, 
            yticklabel style={font=\scriptsize}, 
        xlabel near ticks,
        xtick align=inside,
        ]
        \addplot  [pattern=north east lines, pattern color=orange] 
            table[col sep=comma, x=x, y=id_y] {webvid1M-q10k-q-gt-dist.dat};
        
        \addplot [pattern=north west lines, pattern color=blue] 
            table[col sep=comma, x=x, y=ood_y] {webvid1M-q10k-q-gt-dist.dat};
        \legend{ID Query, OOD Query};
        \end{axis}
    \end{tikzpicture}
    \label{fig:webvid1M-q-gt-dist}
    \text{\small Distance between $1^{st}$NN and query}
\caption{Distances between nearest neighbor to ID (visual)/OOD (textual) queries ($10^4$ queries for each category).}
\label{fig:q-gt-dist-hist}
\end{figure}

\subsubsection{Limitations of Previous Solution for OOD-ANNS}
\label{sec:section-223-ood-diskann} The pioneering graph index designed to tackle the OOD-ANNS problem is 
RobustVamana introduced in OOD-DiskANN \cite{DBLP:journals/corr/abs-2211-12850}. The primary objective of RobustVamana is to use query vectors to add edges in the Vamana graph \cite{jayaram2019diskann}. After linking the base data, queries are also inserted into the Vamana graph. Then, it launches an interconnecting procedure named \textit{RobustStitch} to create a full connection among the closest nodes associated with inserted queries.

\Cref{fig:ood-diskann-improvement} compares the performance of RobustVamana to its original design, Vamana. Searching with OOD workloads, RobustVamana offers 13\% $\sim$ 67\% improvements over Vamana when reaching recall@10=0.9, though it becomes marginal with the increasing recall. However, \finalred{OOD queries on RobustVamana are, on average, 3.9 times, 5.3 times, and 10.0 times slower than ID queries on Vamana for the three datasets.}

The results underscore the inefficiency in OOD-ANNS and highlight \finalred{the substantial potential to design a novel index for improvement in performance concerning OOD workloads.}

\section{Analysis of OOD Workloads in Cross-modal ANNS}
\label{sec:section-3}

In this section, we analyze out-of-distribution workloads to gain insights into the reasons existing approaches are ineffective in achieving high performance in cross-modal ANNS.

\subsection{Underlying Key Differences}

Through in-depth experiments and analysis, we find there are two critical differences between out-of-distribution ANNS (OOD-ANNS) and in-distribution ANNS (ID-ANNS).






\begin{figure}[t]
    \centering
    \includegraphics[width=\linewidth]{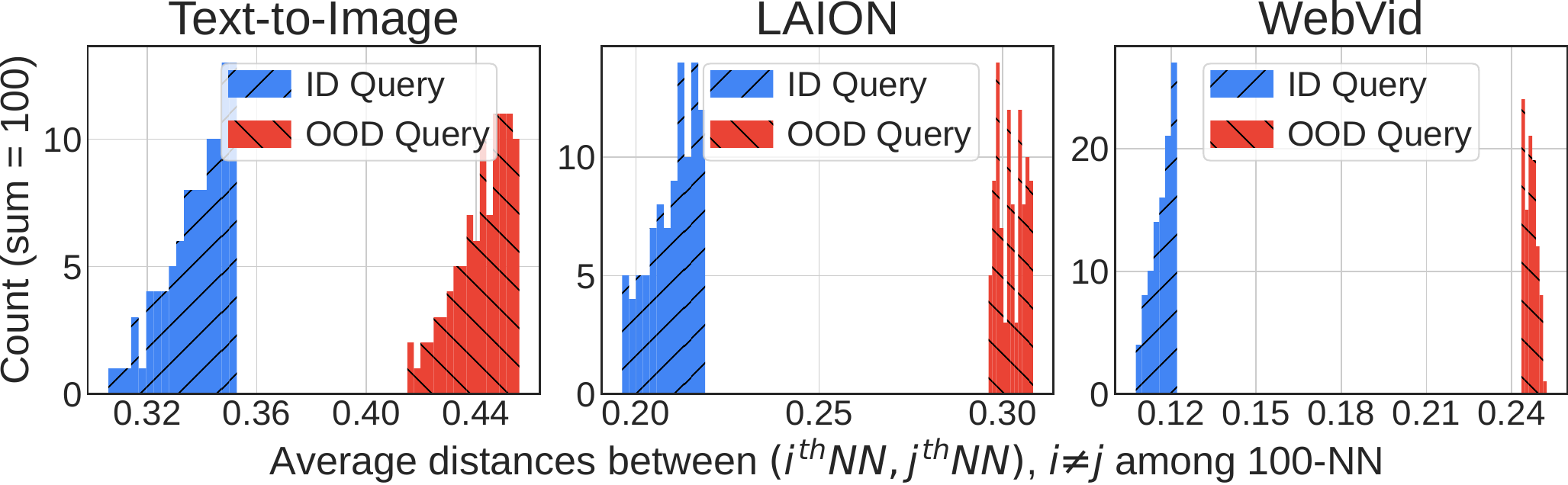}

\caption{For $10^4$ queries' 100 nearest neighbors, the average distances between one vector and the other 99 vectors are computed. Each count represents the mean value of distances within either OOD or ID query sets.}
    \label{fig:multual-hist}%
\end{figure}%

\textbf{OOD queries are distant from \finalred{their nearest neighbors}.}
In \Cref{fig:q-gt-dist-hist}, we calculate the distances between queries and their k-nearest neighbors (k=1), denoted as $\delta(q,i^{th}NN)$. Let $q$ denote the query, and $\delta(\cdot,\cdot)$ represents the distance measurement function of each dataset. As depicted in this figure, we observe that the distances from OOD queries to their nearest neighbors are significantly greater than those from in-distribution queries to the ground truth of the same base data, i.e. $\delta(q_{ood},i^{th}NN_{ood})\gg \delta(q_{id},i^{th}NN_{id})$ in the context of ANNS, with only a small intersection observed in the histograms of Text-to-Image. Despite the presence of extreme values, focusing on the median, OOD queries are 2.1 times, 5.3 times, and 11.3 times farther from their nearest neighbors than ID queries.%
We also find that the deviated OOD query leads to \textbf{the k-nearest neighbors of an OOD query are distant from each other in the high-dimensional space,} as opposed to the neighboring nature of the nearest neighbors of an ID query. To confirm that the k-nearest neighbors of a given OOD query exhibit considerable spatial separation, in the scenario with k=100 for a given query $q$, we calculate the average distance between $i^{th}NN$ ($i=1...100$) and the other 99 nearest neighbors of $q$, resulting in 100 values that represent the degree of separation between nearest neighbors of $q$. Subsequently, the values for $i^{th}NN$ across all queries are averaged to obtain the mean, reflecting this general property. \Cref{fig:multual-hist} presents the phenomenon, distances in neighbors of OOD queries are evidently larger than the neighbors of ID queries, for about 1.29 times, 1.45 times, and 2.11 times \finalred{on the three datasets, respectively}. The finding suggests the presence of numerous noise data vectors between the top-k answers to OOD queries.

\begin{figure}[!t]
    \centering
    \includegraphics[width=0.8\linewidth]{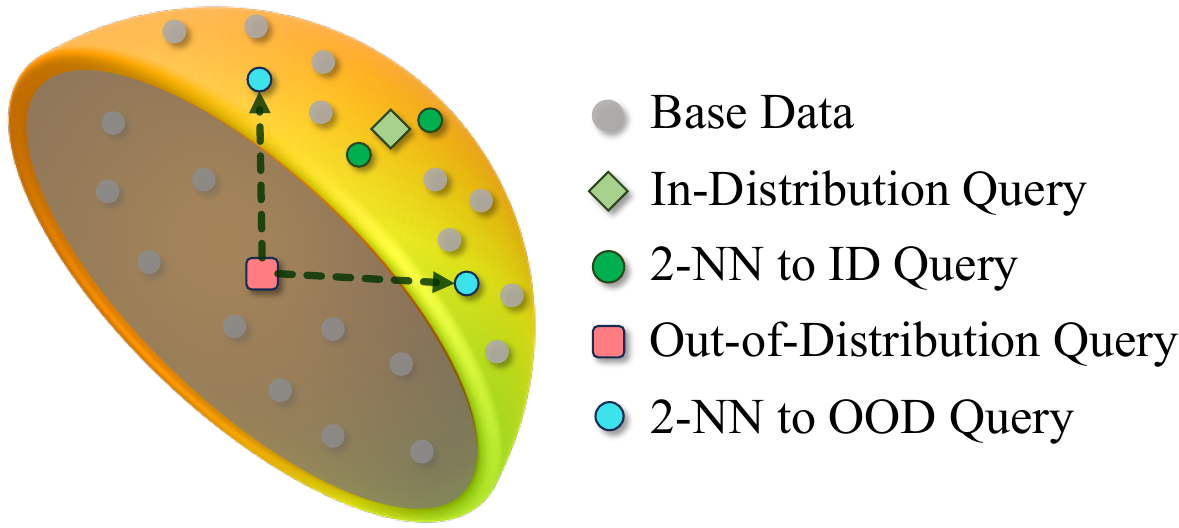}
\caption{A three-dimensional toy example illustrating the challenge of out-of-distribution nearest neighbors search.}
    \label{fig:ood-toy-example}
\end{figure}

To illustrate the counter-intuitive phenomenon, we present a toy example in \Cref{fig:ood-toy-example}. The base data, depicted as grey dots, fluctuates in the vicinity of a 3D spherical surface. An ID query (lime diamond) is situated in close proximity to the surface and is surrounded by two closely located nearest neighbors (green points). Conversely, an OOD query (red square) is positioned far away from the surface of the ball, residing within a section of the half-sphere. The two blue points, spatially distant from each other, sink deeper into the sphere, becoming the closest neighbors for the OOD query.

Real-world examples from the LAION dataset are presented in \Cref{fig:voronoi-example}. This diagram, derived from a subset of the LAION dataset containing 500 data points, is generated by clustering 40 centers from the \finalred{K-means} algorithm. Three examples from both ID queries and OOD queries are sampled, and the five nearest neighbors to each sampled query are computed. Principal Component Analysis (PCA) \cite{hotelling1933analysis} is applied on 100K points to reduce the dimensionality of data to 2-dimension for visualization, \finalred{and sampled 500 vectors are used} to calculate the Voronoi cells. The proximity of the five nearest neighbors to \finalred{ID queries} is evident by the relatively close Voronoi cells that contain them. Conversely, the five nearest neighbors to \finalred{OOD queries} show significant dispersion across the diagram, with the 5-NN of the OOD query residing in separated Voronoi cells.

\subsection{Why Previous Methods Fail on OOD-ANNS}
\label{sec:section-32}





The main reason for the inefficiency of existing approaches is that the OOD query breaks the assumptions held by conventional ANNS approaches. The primary assumptions behind state-of-the-art ANNS indexes include 1) queries exhibit proximity to vectors in the base data, assuming a shared distribution for both queries and base data \cite{DBLP:journals/pvldb/FuXWC19,DBLP:journals/tkde/LiZSWLZL20,DBLP:journals/pacmmod/PengCCYX23}, and 2) the k-nearest neighbors of a query are close to each other in space $\mathbb{R}^{D}$, or \textit{a neighbor’s neighbor is likely also to be a neighbor} \cite{DBLP:journals/pvldb/FuXWC19,DBLP:journals/pacmmod/PengCCYX23,DBLP:journals/tkde/LiZSWLZL20,dong2011efficient,DBLP:conf/icml/ProkhorenkovaS20,shrivastava2023theoretical}. 

Based on the assumptions, the graph-based methods generally utilize beam search both for constructing an approximate k-nearest neighbor graph and executing searches \cite{DBLP:conf/sigmod/LiZAH20,DBLP:journals/pami/FuWC22,DBLP:journals/pvldb/FuXWC19}. Thus, these methods transform spatial proximity vectors into nodes that are closely connected in a graph and further presume that beam search can efficiently navigate into a sphere containing ground truths by shrinking the search space at each step in greedy routing \cite{DBLP:conf/icml/ProkhorenkovaS20,DBLP:journals/pacmmod/PengCCYX23,DBLP:journals/pvldb/FuXWC19}.

However, the search space expands significantly for OOD queries. Considering a high-dimensional sphere denoted as $B^{k}(1^{st}NN, R)$ centered at $1^{st}NN$ of a given query, with the radius $R$ defined as the maximum of $\delta(i^{th}NN, j^{th}NN)$ for a given query ($i \ne j$). The data nodes inside a sphere $B^{s}(x)$ enclosing the currently visiting node $x$ and $B^{k}$ constitute the recognized search space on the graph \cite{DBLP:journals/pvldb/FuXWC19,DBLP:journals/pacmmod/PengCCYX23}. \finalred{In \Cref{fig:multual-hist}, it is observed that $R_{ood}$ is significantly larger than $R_{id}$}, ranging from 1.29 $\times$ to 2.11 $\times$. Expressing the volume of a sphere as $C_{B} \times R^{D}$, where $C_B$ is a constant for a fixed dimension $D$, the ratio $R_{ood}^{D}/R_{id}^{D}$ \finalred{increases substantially in the high-dimensional space}, leading to significant enlargement of $B^{k}_{ood}$ compared to $B^{k}_{id}$. Meanwhile, the search space $B^{s}(x)$ undergoes a vast expansion and encounters challenges in efficiently shrinking due to the inflation of $B^{k}$. Each of the separated nearest neighbors to an OOD query becomes a local optimum trap in greedy routing, posing difficulties for search convergence on existing graph indexes. This indicates that a massive number of nodes are required to be visited for an OOD query.

\begin{figure}[!t]
    \centering
    \includegraphics[width=\linewidth]{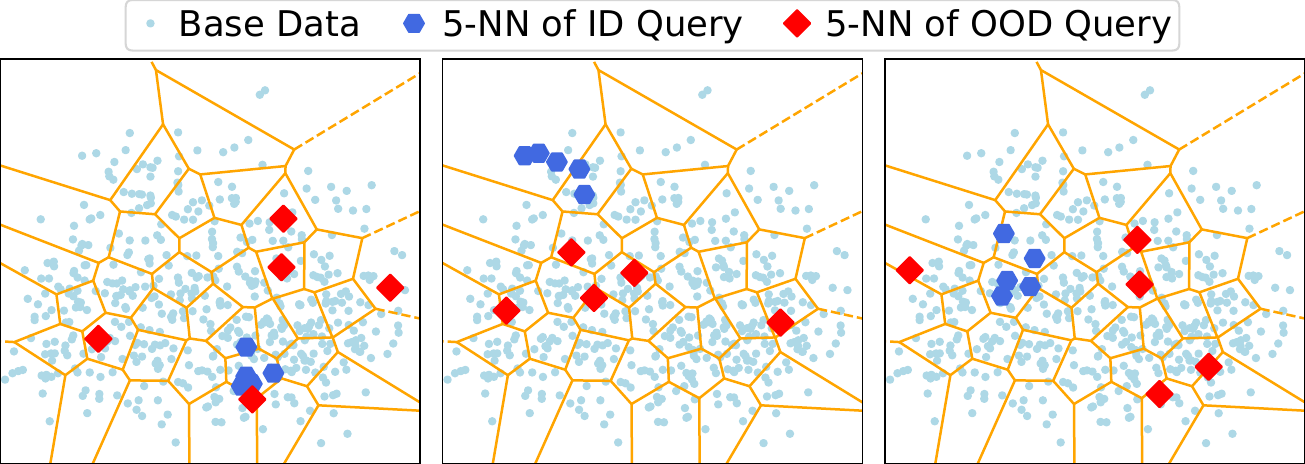}
\caption{Voronoi diagrams generated from 500 base data vectors sampled from the LAION dataset.}
    \label{fig:voronoi-example}
\end{figure}

To illustrate \finalred{the difficulty of convergence when searching an OOD query}, assuming a prevailing graph index is constructed based on the data in \Cref{fig:voronoi-example}, where the close vectors tend to be connected within the graph index. For an ID query, $B^{s}(x)$ is easily \finalred{reduced} along the search path due to the close proximity of the nearest neighbors (blue hexagons). With each progression in greedy routing, the search space contracts to a small sphere containing the nearest neighbors to the ID query \cite{DBLP:journals/pvldb/FuXWC19,DBLP:conf/icml/ProkhorenkovaS20,DBLP:journals/pacmmod/PengCCYX23}. However, for an OOD query, beam search struggles to converge in a single direction. The search space cannot be recognized as shrunk even when routed to one of the k-nearest neighbors (red diamonds). This is because the ground truths for an OOD query are distributed within a huge sphere compared to an ID query. \finalred{To achieve a high recall}, the search process for OOD queries requires a larger beam width, an extended search path, and increased computations and memory accesses along the detour to escape from the local optimum and find dispersed answers, leading to performance degradation.

The space partition-based methods such as IVF also suffer from OOD queries. Clusters are obtained by running \finalred{K-means} algorithm on the base data, where close points are assigned to the same cluster, and the nearest neighbors for OOD queries can be distributed in separated far-away clusters. A cluster containing the closest centroid to an OOD query may not contain \finalred{ground truths} since the \textit{close neighbors} recognized by an OOD query can be dispersed. In the real-world example \Cref{fig:voronoi-example}, the obvious performance impact for partition-based methods is the necessity to scan 5 clusters for OOD queries to achieve recall@5=1.0, compared to only 2 clusters needed for ID queries, resulting in about 2.5 times lower performance. This negative impact becomes worse for million-scale datasets, as described in \Cref{sec:section-2-3-1-ood-eval}.

\noindent
\begin{figure*}[t]
    \centering
    \centering
        
    \includegraphics[width=0.97\textwidth]{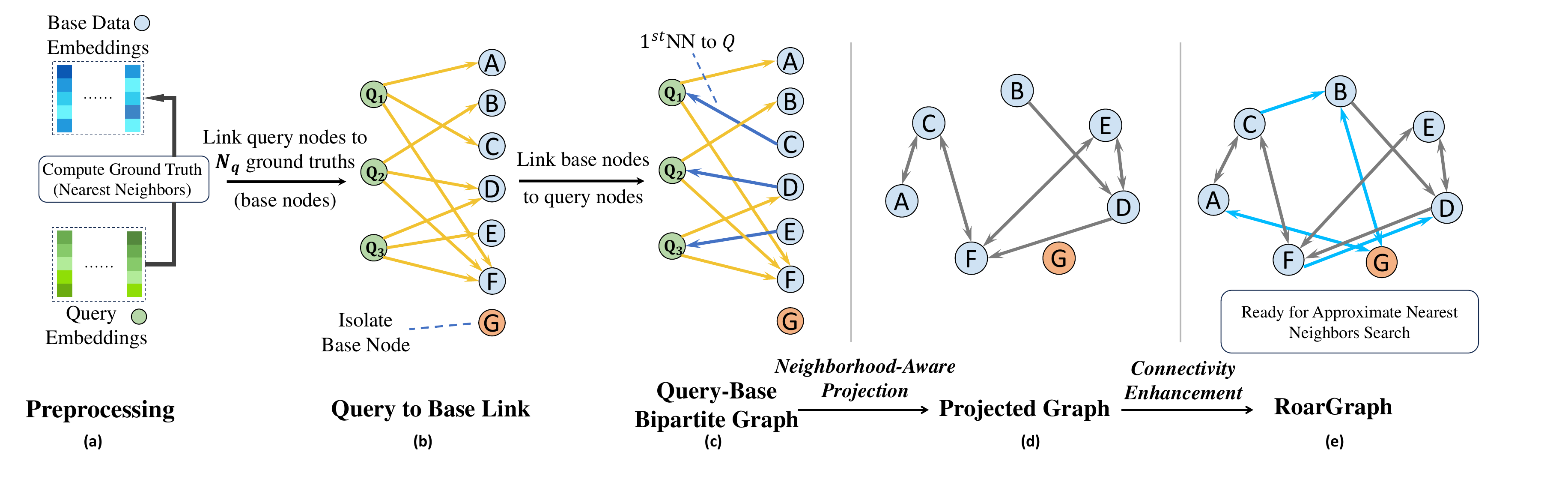}
    
\caption{\revisionred{An overview of RoarGraph construction.}}
    \label{fig:mehod-framework}
\end{figure*}
        
    

\section{\graphname: A graph index for efficient OOD-ANNS}
\label{sec:section-4}

With the revelation of inefficiencies in OOD-ANNS and insightful analyses, we introduce \graphname, a graph index that is built under the guidance of query distribution to provide efficient ANNS in cross-modal retrieval.

\subsection{Query Guided Index for ANNS: Challenges}





\begin{figure}[t]
    \centering
    \includegraphics[width=\linewidth]{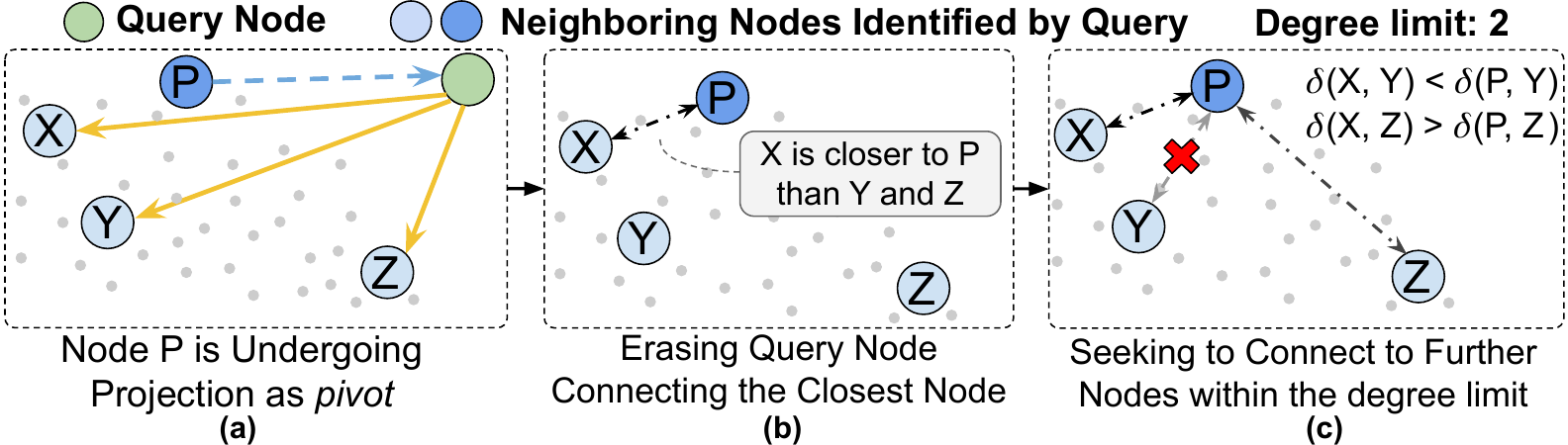}
    
    \caption{\revisionred{An example of \textit{Neighborhood-Aware Projection}. Node P, as the pivot, selects neighbors in the projection.}}
    \label{fig:projection-example}
\end{figure}
As the particular characteristics of the OOD workloads break the assumptions made in the design of existing ANNS approaches, we propose to leverage the query distribution to guide the construction of a graph index. In contrast to connecting base nodes with smaller distances, we propose to \textbf{transform spatially distant vectors, perceived as close from queries' perspectives, into closely connected neighboring nodes within the graph index.} \revisionred{Given the goal, our approach \finalred{utilizes queries to guide} the \graphname construction. For example, we use the embedded vectors of image captions sampled from the original billion-scale LAION dataset to build \graphname supporting text-image search for LAION.}




\label{sec:section-4-1}
To effectively utilize the query distribution in guiding the graph construction, a bipartite graph \revisionred{emerges as} a reasonable candidate for modeling the relationship of closeness \revisionred{between embeddings from two modalities}.
\revisionred{Widely utilized in recommendation systems \cite{wu2022graph,fan2022graph,gurukar2022multibisage}, a bipartite graph comprises two sets of nodes with edges connecting nodes of each set, while nodes within the same set remain unconnected.}

In OOD-ANNS, base data and queries can be treated as two distinct sets (base nodes and query nodes) within a bipartite graph. \revisionred{Here, query nodes play a crucial role in serving} as bridges to \revisionred{identify and connect} the nearest neighbors in base data \revisionred{to query vectors}. \revisionred{Besides, \revisionred{a greedy search on such a bipartite graph can be executed by visiting neighbors' neighbors.} Given an online query, starting from a base node, moving to its out-neighbors (queries) to check out-neighbors (base nodes) of these queries, and then selecting a closer base node to proceed.} While it seems to be a natural structure to model the closeness relationship of two distributions and support search, the following challenges remain to build a competent graph index for efficient ANNS.

\begin{enumerate}
    \item Effective establishment of edges between base nodes and query nodes in the bipartite graph for modeling the neighboring relationship between two modalities and enabling practical greedy routing in ANNS is a non-trivial task. Naively interconnecting nearest neighbors for both types of nodes can only cause excessively high degrees, harming search efficiency.
    \item Searching on a bipartite graph entails traversing through query nodes, and maintaining high out-degrees (e.g., $\ge100$) of query nodes is essential for base data coverage. This leads to heavy overheads at each hop and an increased number of visited nodes along the routing path, slowing the search. Because each base node visiting involves fetching vector data from the main memory to the CPU and calculating its distance to the query, both of which are costly in the node-visiting operation \cite{coleman2022graph}. The more nodes are visited, the more substantial memory access and computational burden are imposed. Therefore, \finalred{node degrees need to be further reduced} while preserving navigable paths for search routing. Additionally, the memory consumption of a bipartite graph is increased due to the inclusion of two node types and edges.
    \item Limitations of reachability and connectivity among base nodes arise when utilizing queries to guide the graph construction. \finalred{Solely depending on the out-neighbors of query nodes in the bipartite graph to cover the entire base data becomes challenging, resulting in isolated nodes or components and impacting search efficiency.}
\end{enumerate}
\subsection{Design and Implementation}
\begin{algorithm}[t]
\SetAlgoLined
\SetNoFillComment
\LinesNumbered
\KwIn{Base data set $\mathcal{X}$, query set $\mathcal{T}$, size of candidate set for neighbor selection  $L$, out-degree bound $M$, $|\mathcal{T}| \times N_q$ closest base nodes for each $t \in \mathcal{T}$}
\KwOut{\graphname Index}
    \tcc{Construct bipartite graph}
    Initialize the bipartite graph $G_{bi} \gets \emptyset$
    
    \For{$t \in \mathcal{T}$} {
    Set $N_q$ closest base nodes to $t$ as $N_{out}(t)$
    
    $x$ $\gets$ closest node to $t$ in $N_{out}(t)$
    
    Add a direct edge from $x$ to $t$
    
    $N_{out}(t) \gets N_{out}(t) \setminus \{x\}$
            
    }
    
    \tcc{Bipartite projection}
    Initialize the projected graph $G_{pj} \gets \emptyset$
    
    $G_{pj} \gets Neighborhood\text{-}AwareProjection(G_{bi},M,L)$
    
    $G' \gets G_{pj}$
    
    \tcc{Connectivity enhancement}
    \For{$x \in \mathcal{X}$}{
        $C \gets BeamSearch(G_{pj},L)$
        
        $N'_{out}(x) \gets AcquireNeighbors(x, C, M)$

        $\forall p\in N'_{out}(x), N'_{out}(p) \gets AcquireNeighbors(p, N'_{out}(p) \cup \{x\}, M)$
    }
        $\forall x \in \mathcal{X}, N_{out\_pj}(x) \gets N'_{out}(x) \cup N_{out\_pj}(x)$
        
\Return $G_{pj}$
\caption{\graphname Construction}
\label{alg:index-build} 
\end{algorithm}
\subsubsection{Overview}
\graphname is proposed to address the aforementioned challenges and deliver superior performance in cross-modal ANNS. The construction of \graphname can be outlined in three steps, as presented in \Cref{fig:mehod-framework}(c-e).


In the initial step, we utilize the \textit{Query-Base Bipartite Graph} (\Cref{fig:mehod-framework}(c)) to unify queries (query nodes) and base data (base nodes) within the same data structure. \revisionred{Prior to the bipartite graph, ground truths of queries vectors are computed during preprocessing (\Cref{fig:mehod-framework}(a)). After that,} We establish edges from each query to its $N_q$ nearest neighbors and then edges from base nodes to queries are added under a restrictive constraint that only the closest base node of its in-neighbor can link to the query \revisionred{(blue arrow in \Cref{fig:mehod-framework}(c))}. This approach achieves two primary goals: \revisionred{1)} creating a closeness mapping between queries and the base data and \revisionred{2)} reducing out-degrees of base nodes to enhance search efficiency on the bipartite graph, thereby addressing \finalred{challenge 1}.

Secondly, \revisionred{we propose a technique named \textit{Neighborhood-Aware Projection} to project the bipartite graph onto base data effectively. Before projecting}, a degree limitation is imposed on every node. For each query node, we choose a connected base node as the \textit{pivot} to select neighbors from the out-neighbors of the corresponding query node. Following the selection of the closest node to the \textit{pivot}, we select farther nodes than the selected ones iteratively. Through \textit{Neighborhood-Aware Projection}, we \finalred{remove} the query nodes but keep the neighboring relationship obtained from the query distribution in the projected graph (\Cref{fig:mehod-framework}(d)). It lowers the average degree of the graph and makes the projected graph become navigable \cite{DBLP:journals/pvldb/FuXWC19,DBLP:journals/pami/MalkovY20}. \revisionred{Consequently, the number of visited nodes during the search is reduced and thus resolves challenge 2.}

In the final step, we apply \textit{Connectivity Enhancement} to the projected graph to address the challenges of collecting isolated nodes, handling separated graph components, and introducing more alternative paths between nodes (solving challenge 3). We traverse every node in the projected graph using beam search, incorporating proximate nodes as diverse supplementary neighbors to each node with an additional degree budget. This process enhances the connectivity and reachability of the graph, completing the \graphname index construction (\Cref{fig:mehod-framework}(e)).

Subsequently, we will provide a detailed introduction to the design and each technical optimization.



\subsubsection{Query-Base Bipartite Graph}
We build the query-base bipartite graph, which functions as a unified container, \finalred{to establish a neighboring map between the distributions of base data and queries.}
Both queries ($\mathcal{T}$) and base data ($\mathcal{X}$) contribute to the formation of the bipartite graph as two distinct node types: query nodes and base nodes. \finalred{Two kinds of directed edges need to be established: \revisionred{1)} edges from base nodes to query nodes and \revisionred{2)} edges from query nodes to base nodes}.



First of all, to build the bipartite graph that can recognize the proximity of the base data from the view of queries, we establish edges from query nodes to base nodes. We add directed edges from each query node to their $N_q$ nearest neighbors (base nodes) in the base data. It is essential to maintain out-degrees of query nodes (\revisionred{$N_q=3$ in \Cref{fig:mehod-framework}(b)}) at a larger value to 1) enlarge coverage of base data and sufficiently model the neighboring relationship within the base data by queries and 2) ensure the overlapping of queries' out-neighbors, making a majority of base nodes within the bipartite graph reachable during search.

Second, \revisionred{to connect base nodes to query nodes, we \finalred{tried} a simplistic strategy that turns existing directed edges into bidirectional ones, assigning base nodes a degree $d=N_{q}$. However, this approach would require checking neighbors' neighbors ($N_{q}^{2}$ nodes) at each step of the search in the bipartite graph, as explained in \Cref{sec:section-4-1}, which is inefficient}. Instead, we propose maintaining $N_q$ links from each query node to its nearest neighbors (see Algorithm 1, line 3) and reducing $d$, making the process more practical. Aligning with our goal to minimize $d$ and adhere to the design goal of modeling closeness relationship, we choose $x$ as the nearest base node among $N_q$ out-neighbors for each query node and add an edge from $x$ to its corresponding query node. This strategy reduces $d$ to 1 and forms the bipartite graph. Concurrently, we remove the link from the query node to $x$, i.e., \(t_c \rightarrow x\) in Algorithm 1, lines 4-6.

\begin{algorithm}[!t]
\SetAlgoLined
\LinesNumbered
\SetNoFillComment
\KwIn{Bipartite graph $G_{bi}=(\mathcal{X},\mathcal{T},E)$, degree limitation $M$, candidate set size $L$, distance function $\delta(\cdot,\cdot)$}
\KwOut{Projected graph $G'$}
$G' \gets \emptyset$

\For{$x \in \mathcal{X}$ and $N_{out}(x) \ne \emptyset$}{
    $S \gets$ $N_{out}(x)$ in $G_{bi}$
    
    $Candidates \gets$ $\emptyset $
    
    
    $\forall s\in S$, add $N_{out}(s)$ in $G_{bi}$ to $Candidates$ until $|Candidates| \ge L$
        
    
    Sort $c\in Candidates$ in ascending order by $\delta(x,c)$
    

    $N'_{out}(x) \gets$ $AcquireNeighbors(x, Candidates, M)$

    \For{$p \in N'_{out}(x)$} {
        
        $N'_{out}(p) \gets AcquireNeighbors(p, N'_{out}(p) \cup \{x\}, M)$
    }
    
}
\Return $G'=(\mathcal{X},E')$
\caption{Neighborhood-Aware Projection}
\label{alg:projection}
\end{algorithm}

\begin{algorithm}[!t]

\SetAlgoLined
\LinesNumbered
\SetNoFillComment
\KwIn{Node $x$, candidate set $C$, degree limitation $M$}
\KwOut{Out-neighbors of $x$}
    $Res \gets \emptyset$ \tcp{$Res$ stores the neighbors for $x$}
    Add the closest node to $x$ in $C$ to $Res$
    
    \For{$c \in C$} {
        $\forall p\in Res$, add $c$ to $Res$ if $\delta(x, c) < \delta(c, p)$
        
        Break when $|Res| \ge M$
            
    }
    \While{ongoing projection \textbf{and} $|Res|<M$}{
        Add $\{c| c\in C \setminus Res\}$ to $Res$
    }

\Return $Res$
\caption{AcquireNeighbors}
\label{alg:rngrulecheck}
\end{algorithm}

\Cref{fig:mehod-framework}(b-c) illustrates the query-base bipartite graph construction. In this example, with $N_q=3$, each query node in the built bipartite graph has two out-neighbors, while the out-degrees of the other base nodes are one. \revisionred{\Cref{fig:mehod-framework}(c), \finalred{node C} serves as $x$ in \Cref{alg:index-build} for query node $Q_{1}$}. Node G is isolated due to degree limit and its insufficient closeness to query nodes, a phenomenon commonly observed in real-world datasets \cite{yandextexttoimage,Bain21,schuhmann2021laion}.

\subsubsection{Neighborhood-Aware Projection}

Despite the bipartite graph's high memory consumption, \revisionred{we find that searching on the query-base bipartite graph is inefficient because routing through query nodes needs a long search path, and there are too many nodes visited along the search path (about $N_{q}$ base nodes at each hop)}. To address the challenge, we propose to project the bipartite graph onto the base nodes. Although a naive bipartite graph projection approach that fully connects nodes sharing common neighbors \cite{zhou2007bipartite} could exclude query nodes, it is undesirable for a graph index due to failing the objective of reducing degrees.
Therefore, we propose \textit{Neighborhood-Aware Projection} to eliminate query nodes from the bipartite graph while preserving the neighborhood relationships of base nodes identified by the query nodes. In \Cref{fig:mehod-framework}(d), the graph is projected, and the out-neighbors of query nodes are linked with \textit{pivots} C, D, and E. The edges $D\rightarrow B$ and $F\rightarrow D$ are not established due to degree bound.

We illustrate the projection with \Cref{fig:projection-example}. Let the query node function as a \textit{bridge} (green node), with the incoming neighbor of a query node designated as the \textit{pivot} (Node P) responsible for selecting its neighbors during projection. Numerous proximal yet irrelevant grey nodes are filtered in the current projection. For each \textit{pivot}, out-neighbors of its \textit{bridges} become potential neighbors of the \textit{pivot} (note that neighbors' neighbors of base nodes are also base nodes). These potential candidates, representing the nearest neighbors to a query but distanced from each other (demonstrated in \Cref{sec:section-3}), are placed into a $Candidates$ queue with a capacity of $L$ and ranked by distances to the pivot (\Cref{alg:projection} lines 5-6).

Next, the closest node in the queue is selected as an out-neighbor for the \textit{pivot} (\Cref{alg:rngrulecheck} line 2, \Cref{fig:projection-example}(b)), followed by acquiring up to degree limit ($M$) neighbors from the $Candidates$ queue for each \textit{pivot}. The essence of acquiring neighbors is that a candidate is excluded from the pivot's out-neighbor list if it is closer to any existing neighbor than to the \textit{pivot} (\Cref{alg:rngrulecheck} line 4). This strategy includes more distant nodes among candidates with the goal of establishing pathways for spatially scattered base nodes that are in close proximity from the queries' perspective. \revisionred{In \Cref{fig:projection-example}(c), we perceive node $Y$ is likely to be reached through node $X$ and $Z$ is harder to find because the $Y$ is \finalred{relatively closer} to $X$, which is already connected to the \textit{pivot} P. So $X$ and $Z$ become neighbors of $P$ within the degree limit.}

To maximize the knowledge from query distribution, we will fulfill out-neighbors within the $M$ degree limitation during projection (\Cref{alg:rngrulecheck} line 8). \revisionred{For example, although node $Y$ was previously filtered, including it in the fulfill operation with degree limit = 3 ensures no degree budget is wasted.} After acquiring neighbors for each pivot, we also check if the pivot can establish reverse links to its in-neighbors (\Cref{alg:projection} line 9).

\begin{figure}[!t]
    \centering
    \includegraphics[width=\linewidth]{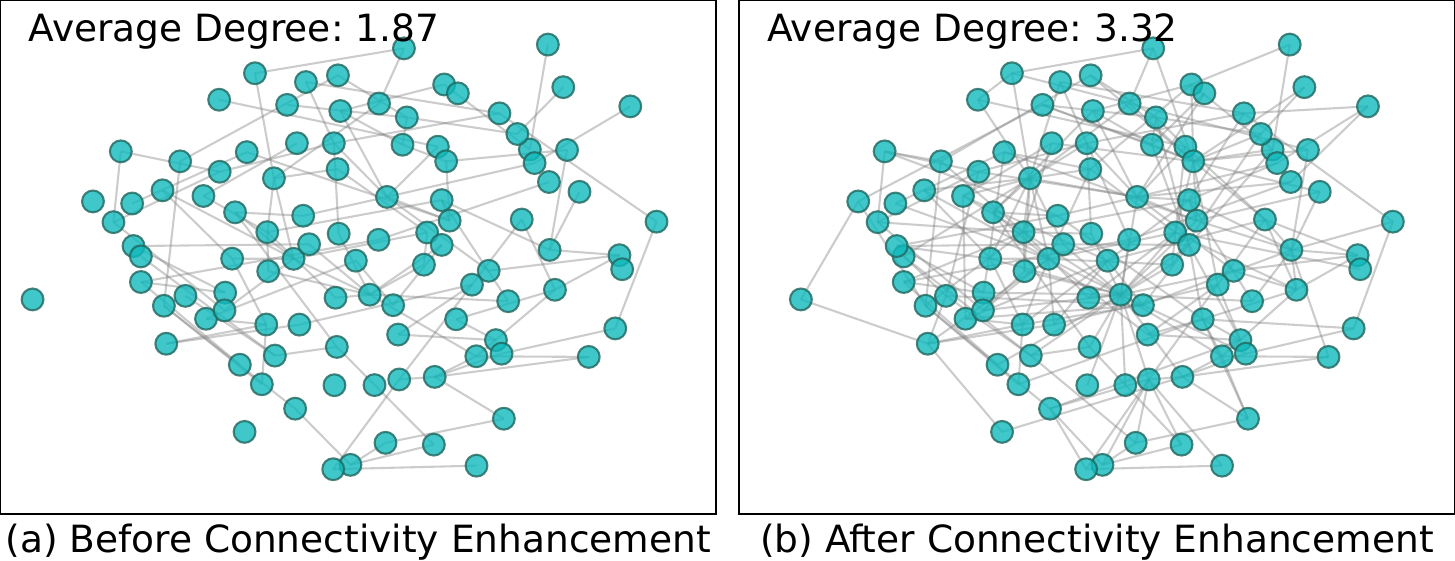}
\caption{An example illustrating connectivity enhancement using 100 vectors of the base data from the LAION dataset.}
    \label{fig:connectivity-example}
\end{figure}





\subsubsection{Connectivity Enhancement}

The projected graph preserves information from the query distribution. Nevertheless, it inadequately provides appropriate reachability and connectivity by only relying on the coverage from the bipartite graph, which is crucial for greedy routing \cite{DBLP:journals/pvldb/FuXWC19,DBLP:conf/icml/ProkhorenkovaS20}. Illustrated in \Cref{fig:mehod-framework}(d), node G is isolated, node B is unreachable, and the path from $A \rightarrow D$ is excessively long. This challenge is further demonstrated in a real-world dataset instance depicted in \Cref{fig:connectivity-example}(a), where \finalred{7\% of nodes are isolated, and 20\% of nodes have degrees less than or equal to one}. This implies that a search process cannot access these 7 nodes even if they are the ground truths, and it may cause a long routing path due to the vulnerable connectivity.

 

To overcome this limitation and enhance navigability, we apply \textit{Connectivity Enhancements} to $G'$, a duplicate of the projected graph (\Cref{alg:index-build} line 10), and then merge edges in $G'$ and $G_{pj}$ in the final step.
Starting from the medoid of the base data, we treat base vectors as queries and conduct a beam search for each of them, with $L$ as the queue capacity. After converging, it returns $L$ visited nodes as results. Every base node gain supplementary neighbors from their individual results through \Cref{alg:rngrulecheck} and also attempt to add reverse edges (\Cref{alg:index-build} line 12-14). Importantly, these supplementary neighbors aid in routing for OOD queries among distant nodes since these edges are established by traversing paths formed from the projected graph.



\Cref{fig:mehod-framework}(c) shows that blue edges \finalred{include} node G into the graph, make node B reachable, and shorten paths between nodes (e.g., A$\rightarrow$D reduced from 4 to 3, B$\rightarrow$A reduced from 4 to 2). \Cref{fig:connectivity-example}(b) shows the effects of connectivity enhancement, which recognized good property to gain navigability for greedy searching, with nodes becoming reachable and links connected for reducing detours \cite{DBLP:journals/pvldb/FuXWC19,DBLP:journals/pami/MalkovY20}. Note that \finalred{the edge direction in this figure is omitted}, and the layout of nodes in this figure does not accurately reflect real geometric relationships; it solely illustrates the graph structure.

\subsection{Search on \graphname}
Greedy routing stands as the conventional search methodology for graph-based ANNS. \graphname, as a general graph index, also employs beam search as the search algorithm. We need to highlight that existing optimizations and compression techniques for ANN indexes \cite{jegou2010product,andre2017accelerated,DBLP:journals/pacmmod/GaoL23,chen2023finger,jegou2010product} are orthogonal with \graphname, so \graphname can directly adopt the optimizations to enhance performance if needed.



In the search process, we utilize the parameter $L$ to control the priority queue length in beam search, with the medoid of the base data as the initial node. At each step, the beam search selects the node $v$ closest to the query from the queue and computes the distances between $v$'s out-neighbors and the query. Then, nodes are added to the queue if they are closer to the query or if the queue is not filled. The search terminates when no closer nodes can be added to the queue.

\section{Experiment}
In this section, we introduce experimental settings, conduct evaluations to compare state-of-the-art graph indexes with \graphname, and present ablation studies.

\subsection{Experimental Setup}
\textbf{Datasets.} Three modern large-scale cross-modal datasets shown in \Cref{table:datasets} are used for evaluation. Text-to-Image \cite{yandextexttoimage} is a popular benchmark with query distribution that differs from the base data, consisting of image and textual query vectors produced by the Se-ResNext-101 model \cite{hu2018squeeze} a variant of the DSSM model \cite{huang2013learning}. The evaluation of similarity relies on the inner product, termed MIPS (maximum inner product search), with a larger value indicating a closer relationship. LAION \cite{radford2021learning} contains millions of Image-Alt-Text pairs used as vector search benchmarks in \cite{tellez2023overview}, the embeddings of image and text are produced by an advanced model, \revisionred{CLIP-ViT-B/32} \cite{DBLP:conf/icml/RadfordKHRGASAM21}. The distance measurement between texts and images is based on cosine distance. WebVid \cite{Bain21} comprises caption and video pairs that are sourced from stock footage sites, with similarity determined by cosine distance. For WebVid, embeddings used in this paper are frame embeddings \revisionred{provided by \cite{clip-video-github}, which are also encoded by CLIP-ViT-B/32.} The official 10K vectors from \cite{yandextexttoimage} are used for querying on Text-to-Image, and two sets of 10K textual vectors sampled from original datasets are used for evaluating LAION and WebVid. In addition, a substantial amount of textual vectors, non-intersecting with queries for evaluation, are provided for index construction within the respective datasets.


\textbf{Algorithms and Parameter Setting.} Graph-based methods are selected as baselines in the evaluations due to their superior performances \cite{DBLP:journals/tkde/LiZSWLZL20}. HNSW \cite{DBLP:journals/pami/MalkovY20} and NSG \cite{DBLP:journals/pvldb/FuXWC19} are widely acknowledged for their search efficiency. The recent $\tau\text{-}$MNG \cite{DBLP:journals/pacmmod/PengCCYX23} establishes additional connections among close nodes, achieving a state-of-the-art performance of recall@100 in ID-ANNS. \revisionred{RobustVamana \cite{DBLP:journals/corr/abs-2211-12850} introduced in \Cref{sec:section-3} is the only index specifically designed for OOD queries.}
We set the best parameters for all algorithms by following the official instructions and empirical experiments.
\begin{itemize}[leftmargin=*]
    \item HNSW \cite{DBLP:journals/pami/MalkovY20}: After varying $M$ from 8 to 48, $M=32$ is set to control the out-degrees of nodes, and $efConstruction=500$.
    \item NSG \cite{DBLP:journals/pacmmod/PengCCYX23}: We set $R=64$ for degree limitation and $C=L=500$ to provide good quality of neighbors.
    \item $\tau\text{-}$MNG \cite{DBLP:journals/pacmmod/PengCCYX23}: $\tau\text{-}$MNG shares the internal parameters $R, C, L$ with NSG, set to 64, 500, 500. The parameter $\tau$ is fine-tuned from 0.01 to 0.3, as suggested in its paper, and for all datasets, $\tau$ is set to 0.01 to achieve the best performance.
    \item RobustVamana \cite{DBLP:journals/corr/abs-2211-12850}: We set $R=64$, $L=500$, and $\alpha=1.0$ after varying it from 1.0 to 1.2. Queries are used in the same quantity as base data for index construction to achieve optimal performance.
    \item \graphname (proposed): $N_{q}=100$ is set to control bipartite graph connections, $M=35$ and $L=500$ is set during \textit{Neighborhood-A
    ware Projection} and \textit{Connectivity Enhancement}. Similar to RobustVamana, queries with the same scale as the base data are used to build indexes.
\end{itemize}

Official codes for all algorithms are employed to build indexes and perform searches. \graphname is implemented in C++, and all source codes were compiled using GCC 10.5.0 with the -O3 optimization.


\textbf{Performance Metrics.} Following previous works \cite{DBLP:journals/pami/FuWC22, DBLP:journals/pvldb/FuXWC19, DBLP:journals/pami/MalkovY20,DBLP:journals/pvldb/LuKXI21,DBLP:journals/pvldb/WangXY021}, we use recall@k (defined in \Cref{sec:section-2-1}), to measure the accuracy of retrieval, and the average recall@k of all queries is reported in evaluations. We adopt queries per second (QPS), which is also equivalent to latency in the single-thread scenario, to measure the search speed, following \cite{DBLP:journals/pvldb/WangXY021,DBLP:journals/pami/FuWC22,DBLP:journals/pvldb/FuXWC19,DBLP:journals/pami/MalkovY20,DBLP:journals/pacmmod/GaoL23,DBLP:journals/pacmmod/PengCCYX23}. To thoroughly validate the efficacy of each method across diverse retrieval scenarios, we configure $k$ for recall@k as 1, 10, and 100 during the evaluation.

Evaluations were conducted on a machine with dual Intel(R) Xeon(R) Gold 5318Y CPU and 512 GB of memory, running in Ubuntu 20.04. To ensure a fair comparison, all algorithms were run in single-thread mode since not all methods support multi-threading in their official implementations.

\begin{figure}[t]
    \centering
    \includegraphics[width=\linewidth]{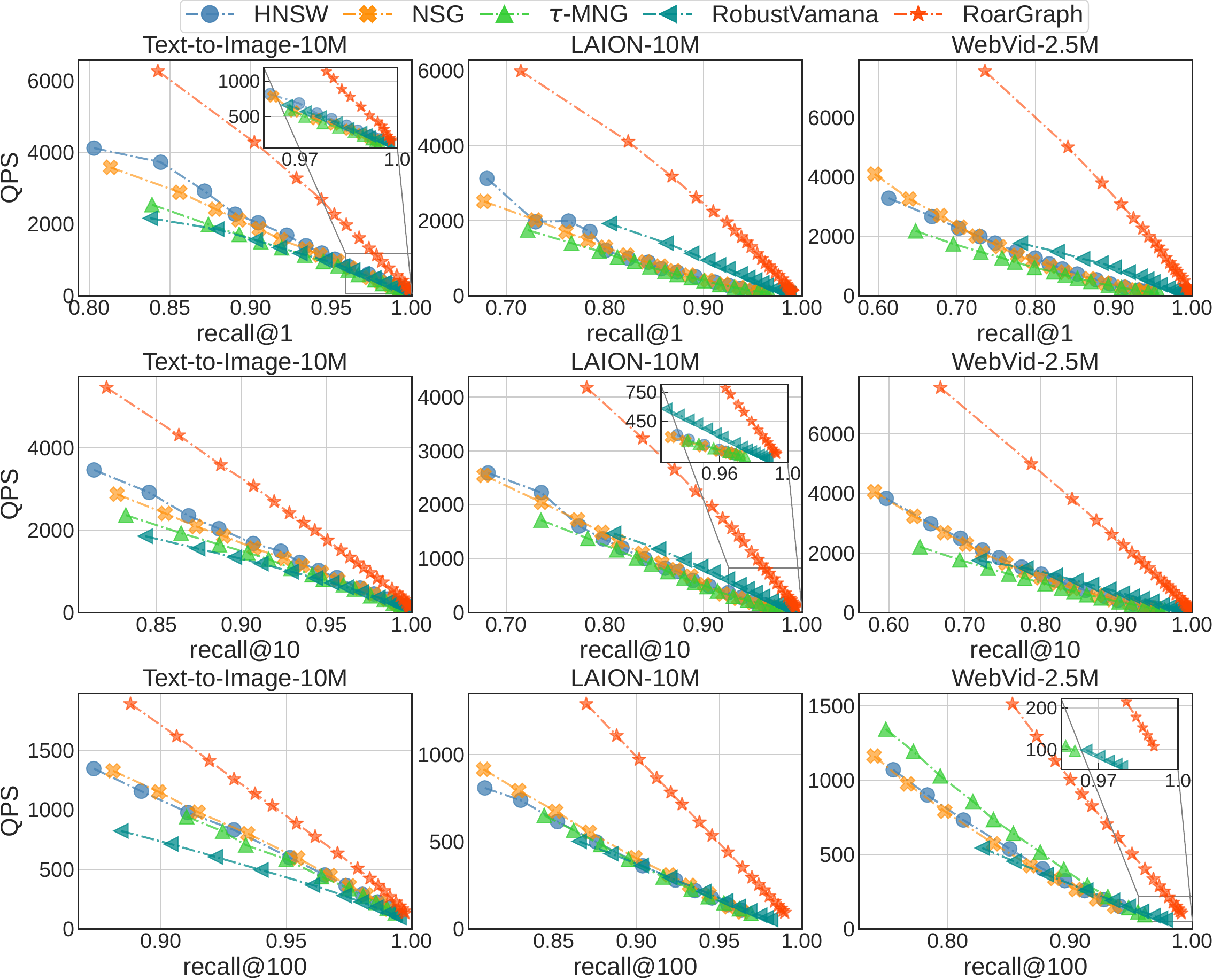}
\caption{Search performance on three datasets. The top right is better.}
    \label{fig:eval-qps-vs-recall}
\end{figure}

\subsection{Search Speed vs. Recall}

Results for QPS vs. Recall are reported in \Cref{fig:eval-qps-vs-recall}. Traditional graph-based methods that are designed for ID search, including HNSW, NSG, and $\tau\text{-}$MNG, work poorly for OOD workloads in cross-modal ANNS. RobustVamana, as a specific solution for OOD tasks, is faster than the three ID indexes on LAION and WebVid for recall@1 and recall@10. The proposed method \graphname consistently outperforms all state-of-the-art graph indexes among three cross-modal datasets when varying $k$ in recall@k from 1 to 100 in all recall regimes. Specifically, for recall@10$\ge$0.9, \graphname achieves speed-ups of 1.84$\times$, 2.58$\times$, and 3.56$\times$ than the most efficient graph index on Text-to-Image, LAION, and WebVid, respectively. Similar improvements are observed for $k=1,100$. Besides, \graphname demonstrates the capability to attain a recall@100 $\ge$ 0.99 or even higher on LAION and WebVid, a level that none of the other graph indexes can practically achieve.

There is an observation that verifies our motivation and analyses in \Cref{sec:section-3}.  For the three conventional graph algorithms, HNSW and NSG have similar performance, while the more recent index $\tau\text{-}$MNG \cite{DBLP:journals/pacmmod/PengCCYX23} shows a lower performance among cross-modal datasets, especially for $k=1\text{ and }10$. This is mainly because $\tau\text{-}$MNG actually adds more edges to connect near neighbors around each node upon NSG. However, according to our insights, the nearest neighbors of an OOD query are not always clustered as ID queries. Ground truths may be widely distributed, so the close nodes connected contribute little to search efficiency but exacerbate the computation burden.

\subsection{Routing Hops vs. Recall}



\begin{figure}[t]
    \centering
    \includegraphics[width=\linewidth]{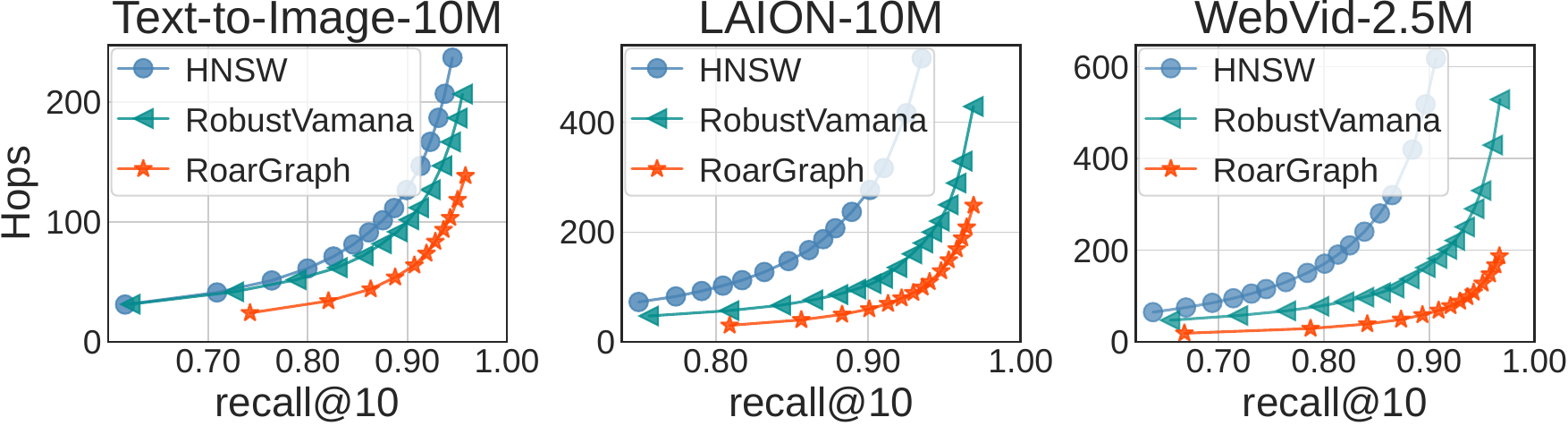}
\caption{Evaluation of hops vs. recall.}
    \label{fig:eval-hops-vs-recall}

\end{figure}

The number of hops required to achieve a target recall serves as an alternative indicator of the search efficiency in navigating the graph-based index. Each hop during the search involves a certain computational cost for checking its neighbors in the graph. A lower number of hops typically signifies a shorter search path when searching on the graph \cite{jayaram2019diskann,DBLP:journals/tkde/LiZSWLZL20}. In \Cref{fig:eval-hops-vs-recall}, we assess the hops incurred during beam search on three representative graph indexes, HNSW, RobustVamana, and \graphname. We observe that \graphname involves much less hops during the search across three datasets compared to HNSW and RobustVamana, reducing the hops (i.e., length of search path) to 44.1\%, 21.0\%, and 10.9\% compared to HNSW, and 53.1\%, 54.7\%, and 41.1\% compared to RobustVamana when recall@10 $\ge$ 0.90. The ratio diminishes with the increasing recall@k, underscoring that \graphname \finalred{established} effective edges for navigating cross-modal queries.


\begin{figure}[t]
    \centering
    \includegraphics[width=\linewidth]{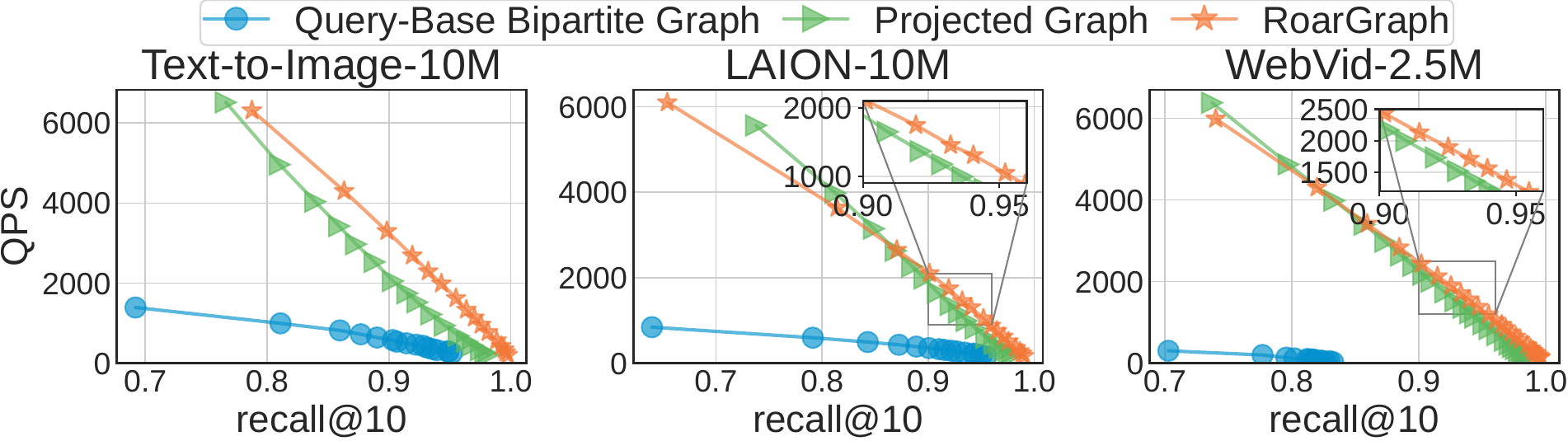}
    \caption{Ablation comparisons.}
    \label{fig:abltion}
\end{figure}

\subsection{Ablation Study}


We perform a comparative evaluation to validate the effectiveness of techniques used in building \graphname. \revisionred{The \textit{Query-Base Bipartite Graph} and the \textit{Projected Graph} created before deriving \graphname are involved, denoted as $G_{bi}$ and $G_{pj}$, both of which are capable of querying. We use QPS-recall in evaluations instead of reporting hops-recall.} \revisionred{This is because, unlike the comparison of HNSW and \graphname that have a similar degree bound, the bipartite graph necessarily maintains a high out-degrees of query nodes. It incurs more nodes to visit at each hop than $G_{pj}$ and \graphname, as a result, the reduction in hop costs can not match the QPS improvement in the comparison of $G_{bi}$, $G_{pj}$, and \graphname. For instance, on the LAION dataset, queries traversing hops $\ge$100 visited 10507, 2426, and 3494 nodes, with recall@10 = 0.918, 0.906, and 0.937 for the Query-Base Bipartite Graph, Projected Graph, and RoarGraph, respectively. These statistics demonstrate that $G_{bi}$ requires visiting more nodes per hop. It also reveals that the Projected Graph benefits from the \textit{Neighborhood-Aware Projection} technique, which effectively decreases node degrees and shows an ability to navigate greedy routing for OOD queries, with only visiting 2426 nodes to achieve recall@10=0.9 in about 100 hops. The \textit{Connectivity Enhancement} provides supplementary edges with increased degrees, improves reachability and connectivity, and we believe it introduces more alternative pathways for search routing, thereby shortening the search path and enhancing search accuracy in a high recall regime.}

The results of performance evaluation in \Cref{fig:abltion} show that the bipartite graph exhibits the lowest performance across the three datasets as analyzed, whereas the Projected Graph achieves multiple times acceleration in all recall regimes through \textit{Neighborhood-Aware
Projection}. We observe that \graphname performs consistently 1.49 $\times$ faster than the Projected Graph on Text-to-Image, but the Projected Graph demonstrates better efficiency when recall@10 $\le$ 0.86 on LAION and WebVid datasets. This observation can be attributed to the increased node degrees after \textit{Connectivity Enhancement}. The supplementary edges introduce additional computation overhead for low recall rates, however, they prove instrumental in finding effective paths for convergence at a high recall level.

\subsection{Effects of Query Set Size for Indexing }
\label{sec:5-5}

To assess RoarGraph's sensitivity to the number of queries ($|\mathcal{T}|$) used in construction, we evaluate QPS-recall tradeoff for different query set sizes by varying the coefficient $p$ in $|\mathcal{T}|=p\times |\mathcal{X}|$ during index construction, the query set is denoted as $\mathcal{T}_p$.

\begin{figure}[t]
    \centering
    \includegraphics[width=\linewidth]{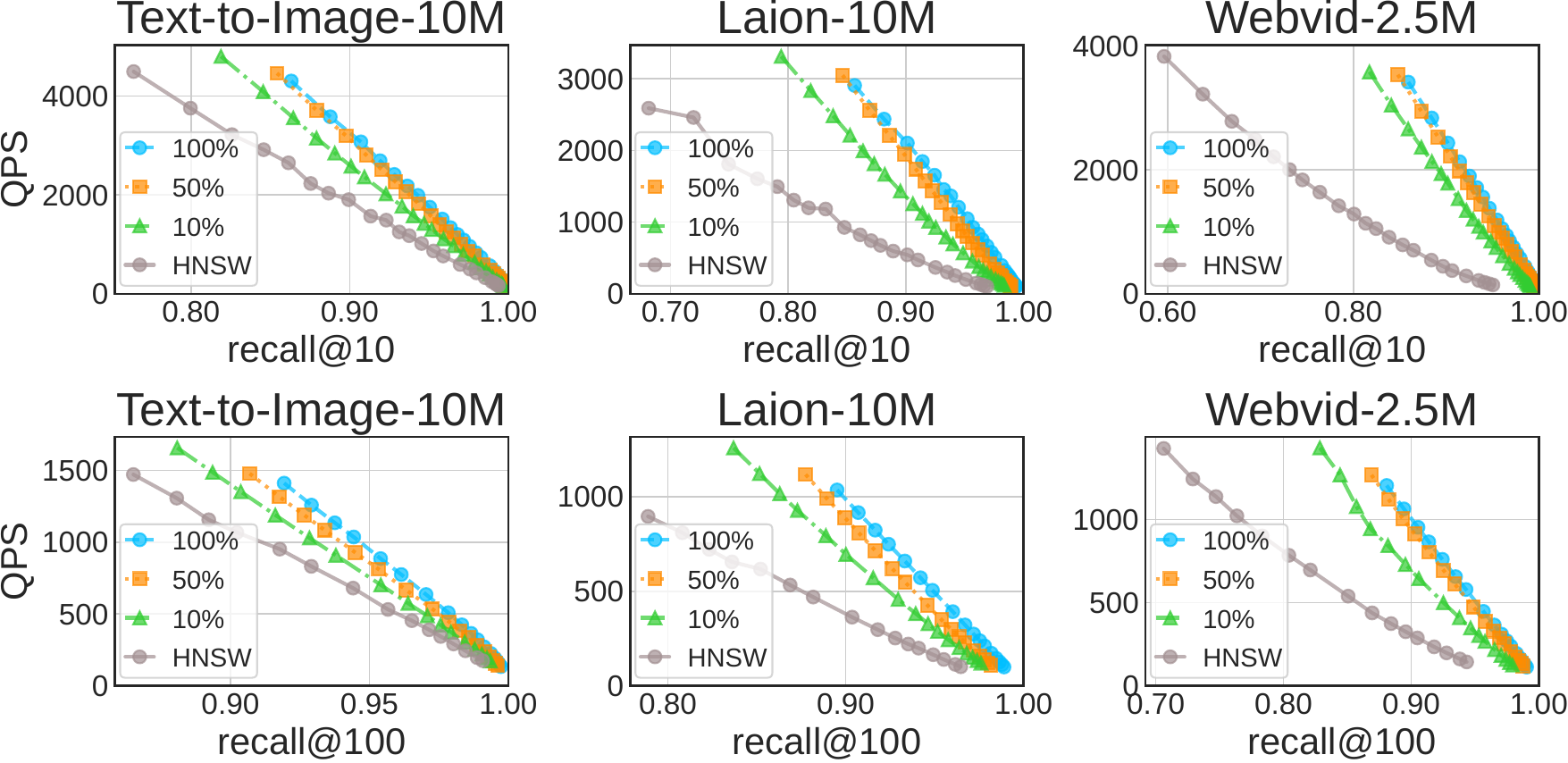}
\caption{Performance comparison across different query set sizes for index construction.}
    \label{fig:diff-train-size}
\end{figure}

\Cref{fig:diff-train-size} displays how search speed correlates with recall rates across different query set sizes during the index construction process, with evaluations conducted at $p=10\%$, $50\%$, and $100\%$ of the base data size. It reveals that employing $\mathcal{T}_{0.5}$ in \graphname construction achieves performance comparable to that built with $\mathcal{T}_{1}$ for both recall@10 and recall@100 metrics in all datasets. Despite a performance decline with reduced query set sizes, indexes constructed from $\mathcal{T}_{0.1}$ are only $11.3\%$, $12.9\%$, and $29.2\%$ \finalred{slower} in reaching recall@100 $\geq0.95$ compared to those constructed with $\mathcal{T}_{0.5}$. Importantly, the results highlight that \graphname maintains superior efficiency, outperforming HNSW by $1.44-4.38\times$ in achieving recall@$10=0.9$ across three datasets, even when only $\mathcal{T}_{0.1}$ are used for building \graphname.




\subsection{Robustness to In-Distribution Query}


In addition to exhibiting superior performance in OOD-ANNS, it is essential that an ANNS index effectively handles in-distribution (ID) queries so that one index can serve different query types in the application. In this experiment, we undertake a comparative analysis utilizing ID queries as the workload for single-modal ANNS evaluation. For each of the three datasets, ID queries comprise 10K visual embeddings sampled from the original large-scale dataset.



\begin{figure}[t]
    \centering
    \includegraphics[width=\linewidth]{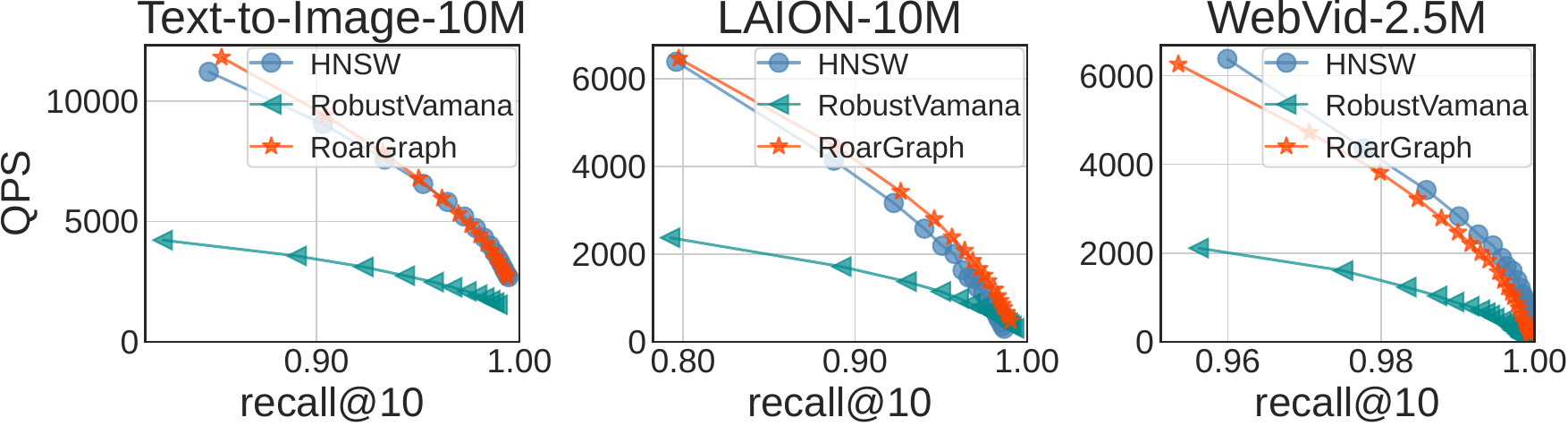}
\caption{Evaluation on in-distribution workloads.}
    \label{fig:id-query-eval}
\end{figure}

As shown in \Cref{fig:id-query-eval}, \graphname demonstrates robustness to ID workloads, offering competitive efficiency against HNSW across three datasets. RobustVamana, although generally slower than \graphname and HNSW, attains recall@10 $\ge$ 0.995 on LAION, where both HNSW and \graphname fail to reach.

\subsection{Index Size and Construction Overhead}

\begin{figure}[t]
    \centering
    \includegraphics[width=\linewidth]{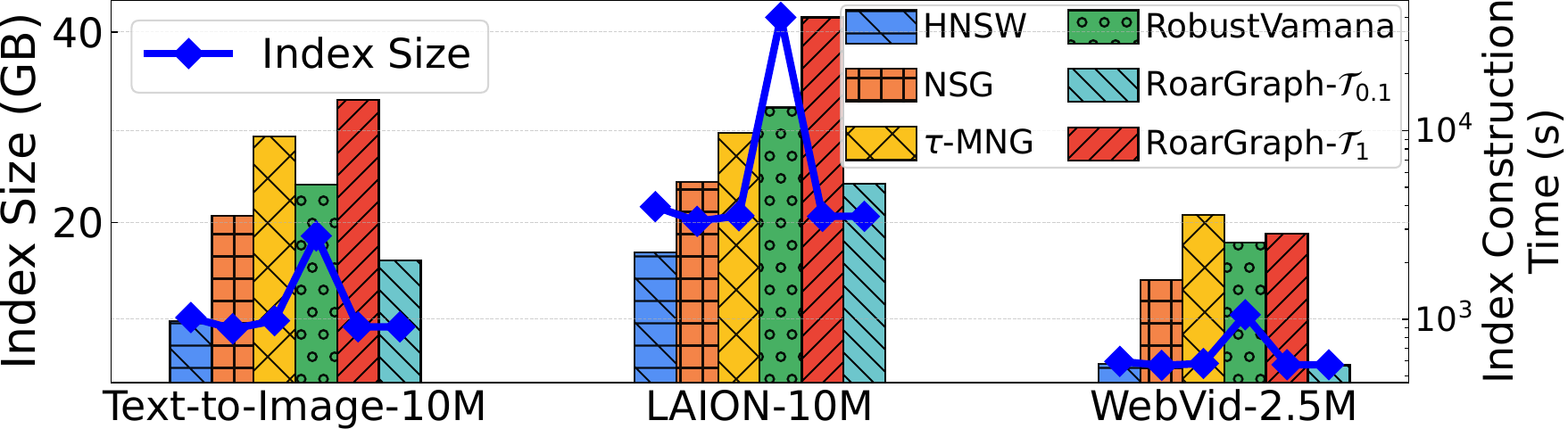}
\caption{Comparison of index sizes and construction overheads. Bars stand for the time costs.}
    \label{fig:index-size-cmp}
\end{figure}

\finalred{\Cref{fig:index-size-cmp} compares} the index size and index build overheads. The final index sizes reflect the memory consumption in the search phase. As shown in the figure, \graphname indexes consume 9.04 GB, 20.64 GB, and 5.07 GB for Text-to-Image, LAION, WebVid, respectively -- only slightly larger than NSG. The result shows that \graphname maintains a memory-friendly small index size but provides significant improvements in cross-modal ANNS.

\revisionred{In the evaluation of index construction overheads, we use 64 threads for all graph indexes. We compared \graphname's construction using two query vector set sizes: 100\% and 10\% ($\mathcal{T}_1$ and $\mathcal{T}_{0.1}$), relative to the base data volume. 
HNSW emerged as the most time-efficient index to build across all datasets. \finalred{Both NSG and $\tau\text{-}MNG$ require to build an approximate nearest neighbor graph that increases the construction time}. Specifically, utilizing $\mathcal{T}_{1}$ in \graphname's construction took 1.12 to 3.02 times longer than RobustVamana, 1.7 to 7.5 times longer than NSG, and 4.8 to 17.5 times longer than HNSW. However, \graphname is 21\% faster compared to \(\tau\)-MNG on WebVid. The preprocessing phase in \graphname, which computes ground truths of query vectors, accounts for 87\% to 93\% of the total construction time with $\mathcal{T}_1$.}
\revisionred{
When constructed with $\mathcal{T}_{0.1}$, \graphname's construction time is significantly reduced, taking only 16\% to 54\% as long as \(\tau\)-MNG, and 35\% to 98\% as long as NSG. While constructing \graphname with $\mathcal{T}_{0.1}$ is still twice as long to construct on the Text-to-Image and LAION datasets, it matches HNSW's build time on WebVid. In this case, the preprocessing phase occupies 67\%, 73\%, and 43\% of the entire index construction time on Text-to-Image, LAION, and WebVid, respectively.}

\revisionred{Considering RoarGraph constructed with $\mathcal{T}_{0.1}$ still offers notable search performance improvement (see \Cref{sec:5-5}) while significantly reducing construction time, this allows applications to make tradeoffs between construction costs and search efficiency, making it a flexible and practical choice for different scenarios.}

\section{Discussion}

\textbf{\graphname construction in real-world scenarios}. \revisionred{The key idea of RoarGraph is to effectively utilize cross-modal query vectors to build a graph index that provides significant improvements in cross-modal vector search performance, and it can be adopted in the following real-world scenarios.}

\revisionred{Vast amounts of historical queries are available in large-scale embedding-based information retrieval and recommendation systems, as evidenced by Bing \cite{tian2023pass}, YouTube \cite{covington2016deep}, Amazon \cite{Yu2024,Lee2024,Chang2024}, TikTok \cite{liu2022monolith,gao2020deep}, Pinterest \cite{gurukar2022multibisage}, etc. Therefore, as a workload-driven cross-modal ANNS index incorporating knowledge from query vectors, RoarGraph can effectively utilize the historical query vectors to build the index. Besides, such applications can also use their multimodal deep-learning models, which produce embeddings for both base data and queries, to encode queries from large-scale (up to billions) public real-world datasets \cite{miech2019howto100m, schuhmann2022laion,Bain21,schuhmann2021laion} if supplementary data is needed.} \finalred{Under situations with a limited number of queries, RoarGraph can still yield notable performance improvements (as described in \Cref{sec:5-5}).}

\flushcolsend
\begin{figure}[t]
    \centering
    \includegraphics[width=\linewidth]{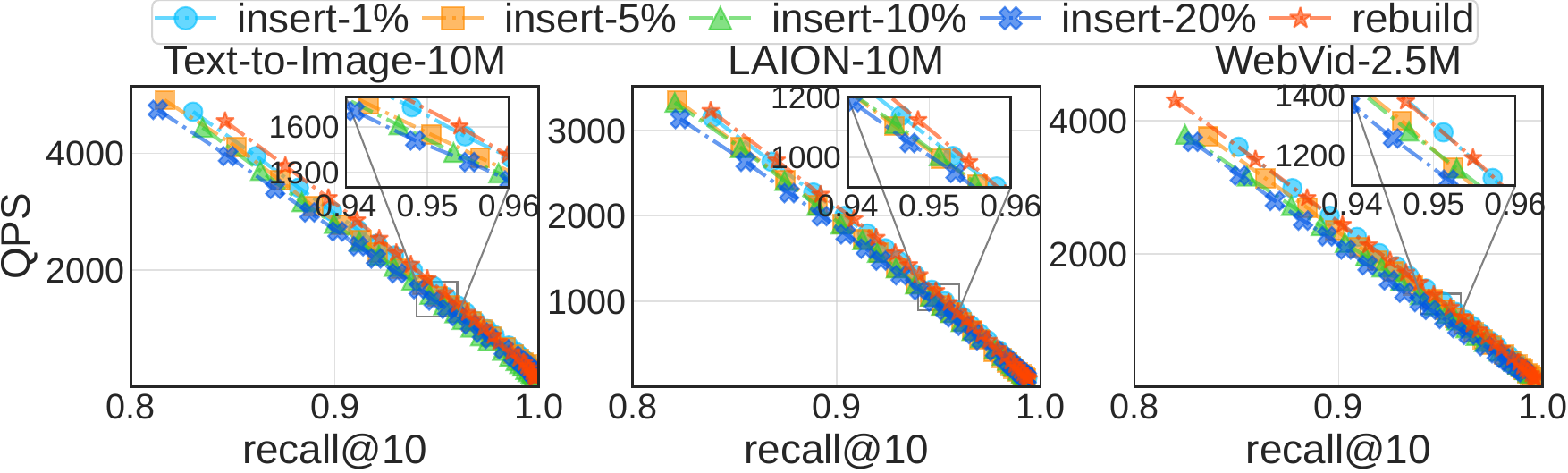}
\caption{Impacts from different amounts of insertions.}
    \label{fig:insertion-eval}
\end{figure}
\balance
\textbf{Update in \graphname.} \revisionred{The query-base bipartite graph during index construction is saved to facilitate offline insertion for \graphname. The insertion begins by considering the incoming base data vector $v$ as a query to search its approximate nearest neighbor within \graphname, and only the base node connected by at least one query node $q$ will be returned as a result. In cases where multiple query nodes are eligible, the nearest $q$ to $v$ is selected. The next step is to leverage the out-neighbors of $q$, denoted as $N_{out}(q)$, to integrate $v$ and $q$ as a sub-bipartite graph ($N_{out}(q)\cup q\cup v$). Following this, \textit{Neighborhood-Aware Projection} is applied on the sub-bipartite graph, with $v$ serving as \textit{pivot}. This produces a projected graph comprising the new vector $v$, and RoarGraph merges newly established edges associated with $v$ to complete the insertion. Additionally, the bipartite graph is updated to include $v$ among the out-neighbors of $q$, paving the way for subsequent insertion.}
\revisionred{The insertion strategy is efficient because we avoid the costly process of computing exact distances between incoming data and query nodes in the bipartite graph. It only takes 583 seconds to insert two million base data vectors from the LAION dataset into a RoarGraph with 64 threads. This time consumption constitutes only 7\% of the time required to rebuild the index.}

\revisionred{We evaluated indexes after different numbers of insertions and compared them to the reconstructed index. The volumes of the inserted data are proportional to the dataset scale, ranging from 1-20\%. \Cref{fig:insertion-eval} shows that RoarGraph indexes updated through this insertion strategy are competitive to those rebuilt for including new data. After handling \finalred{the insertion} of two million vectors into RoarGraphs constructed from Text-to-Image and LAION datasets, RoarGraph shows about 13\% and 10\% performance deterioration relative to the reconstructed index when recall@10=0.95. On \finalred{WebVid-2.5M}, a 17\% QPS reduction relative to the rebuilt index for recall@10=0.95 is observed after inserting 0.5 million vectors.} 

\revisionred{For deletions, RoarGraph adopts tombstones to mark deleted points \cite{DBLP:journals/pacmmod/PengCCYX23,xu2022proximity}. Deleted points participate in routing but are excluded from results.
}

\revisionred{
The current insertion approach faces challenges when handling continuous massive insertions, as the subsequent bipartite graph approximates closeness relationships less precisely than the reconstructed index. Besides, tombstones require periodic reconstruction as updates accumulate \cite{DBLP:journals/pacmmod/PengCCYX23}. These challenges indicate that both insertion and deletion methods need explorations in future work.
}




\section{Related Work}
The literature on ANNS is extensive. Despite the approaches introduced above, there are studies that utilize queries to improve the performance of graph-based ANNS. In the work of \cite{DBLP:conf/sigmod/LiZAH20}, it formulates the termination condition of beam search as a binary classification task, then utilizes queries to train a classification model to determine when to terminate the search. Introducing graph convolutional networks (GCN) into ANNS, \cite{DBLP:conf/icml/BaranchukPSB19} aims to discover an optimal search routing path through learning from extensive training queries. The study conducted by \cite{feng2023reinforcement} combines reinforcement learning with GCN to guide the search routing on the proximity graph. To learn from query distribution and prune edges within a graph index, GraSP \cite{zhang2022grasp} employs a probabilistic model and subgraph sampling to learn the importance score of edges and prune the graph. However, learning-based methods require costly training and tuning phases during index construction.




Using neural networks for similarity ranking in recommendation systems, \cite{tan2021fast} utilizes a bipartite graph to connect users and items and execute the search on the bipartite graph. \finalred{It claims that there is no defined similarity metric within item vectors or user vectors, so edges should be established only between the two types of nodes. This scenario differs from cross-modal retrieval, where the similarity between different base vectors or query vectors can also be evaluated.}

\finalred{ScaNN \cite{guo2020accelerating} uses vector quantization (VQ) \cite{gray1984vector} for partition and Product Quantization (PQ) \cite{jegou2010product} for compression with anisotropic loss. It applies optimizations, such as scalar quantization, rescoring, and SIMD in-register PQ lookup \cite{andre2017accelerated} for fast search.}


Several theoretical analyses of graph-based methods in ANNS have been conducted \cite{DBLP:conf/icml/ProkhorenkovaS20,DBLP:journals/pvldb/FuXWC19,DBLP:journals/pacmmod/PengCCYX23,shrivastava2023theoretical,DBLP:journals/pami/FuWC22}. The time complexity of the search process for the graph index, as derived from these studies, is based on the assumption that the query follows the same distribution as the base data. However, these proofs face challenges when applied to cross-modal ANNS.

\section{Conclusion}
\flushcolsend
As an important workload from cross-modal data retrieval, we perform an insightful analysis of OOD queries in this paper.
We find that the k-nearest neighbors of an OOD query are distant from each other in the \finalred{high-dimensional embedding space. This is the root cause of the inefficiency of existing ANNS approaches, as this characteristic} breaks the assumptions of their designs. We propose \graphname, an efficient graph index for OOD-ANNS, which is constructed under the guidance of query distribution. Extensive experimental results demonstrate the superior performance of \graphname on cross-modal vector search.

\flushcolsend


\bibliographystyle{ACM-Reference-Format}
\bibliography{ann_ref}

\end{document}